\documentclass[numbook, envcountsame, envcountreset]{svjour3}
\usepackage{amssymb}
\usepackage{graphicx}

\newtheorem{prop}[theorem]{Proposition}
 
\newtheorem{exam}[theorem]{Example}

\def\eqref#1{(\ref{#1})}
\newcommand{\eqa}{\begin{eqnarray}}
\newcommand{\eeqa}{\end{eqnarray}}
\newcommand{\beq}{\begin{equation}}
\newcommand{\eeq}{\end{equation}}
\newcommand{\nn}{\nonumber}

\newcommand{\pal}{\partial}

\newcommand{\de}{\delta}

\newcommand{\lm}{\lambda}

\newcommand{\pf}{\noindent{\it Proof \ }}

\newcommand{\res}{{\rm res}}

\newcommand{\cL}{{\mathcal L}}
\newcommand{\epf}{$\quad$\hfill
\raisebox{0.11truecm}{\fbox{}}\par\vskip0.4truecm}

\begin{document}

\title{Infinite-Dimensional Frobenius Manifolds for $2+1$ Integrable Systems}
\subtitle{}

\author{Guido Carlet \and Boris Dubrovin \and Luca Philippe Mertens}

\institute{SISSA-ISAS, Via Beirut 2--4, 34014 Trieste, Italy \\ 
\email{carlet@sissa.it, dubrovin@sissa.it, mertens@sissa.it}}

\maketitle


\begin{abstract}
We introduce a structure of an infinite-dimensional Frobenius manifold on a subspace in the space of pairs of functions analytic inside/outside the unit circle with simple poles at $0/\infty$ respectively. The dispersionless 2D Toda equations are embedded into a bigger integrable hierarchy associated with this Frobenius manifold.
\end{abstract}


\section{Introduction}

\subsection{\bf Frobenius manifolds and integrable hierarchies}\par 
Frobenius manifolds, originally invented \cite{D92} as the geometric set-up for the WDVV equations of two-dimensional topological field theory \cite{wi}, proved to be an efficient tool in studying integrable systems of PDEs with one spatial dimension. For an integrable system of $n$ evolutionary PDEs
\beq\label{hier}
\pal_t u_i =K_i(u; u_x, u_{xx}, \dots; \epsilon), \quad i=1, \dots, n
\eeq
depending on a small parameter $\epsilon$, the structure of the small dispersion limit $\epsilon\to 0$ (if the limit exists) under very general assumptions of existence of a bihamiltonian structure and a tau-function is completely described by a suitable $n$-dimensional Frobenius manifold (see details in \cite{normal}). In particular, a basis of the first integrals of the dispersionless hierarchy can be efficiently computed in terms of the flat sections of the canonical deformed connection, the bihamiltonian structure of the hierarchy is expressed via the natural flat pencil of metrics on the Frobenius manifolds etc. (see \cite{icm} and references therein). The structure of the full hierarchy \eqref{hier} is completely determined by the Frobenius manifold along with $n$ functional parameters (functions of one variable) called {\it central invariants} \cite{cpam}.

Our goal is to extend the above programme to integrable PDEs with two spatial dimensions (the so-called 2+1 systems). In particular we expect that the technique of Frobenius manifolds will be instrumental in the description of a complete set of first integrals of integrable dispersionless systems also for the 2+1  case thus giving a way to studying the properties of general solutions to these systems under very mild analytic assumptions. It might also be helpful in the classification of various dispersive deformations of the dispersionless 2+1 systems according to the scheme of \cite{cpam}.

In the present paper we consider just one example of a 2+1 integrable system, namely, the 2D Toda equation
\beq\label{tod1}
\pal_t^2 u_n -\pal_y^2 u_n = e^{u_{n+1}}-2 e^{u_n}+e^{u_{n-1}}.
\eeq
In this case we have two spatial variables: a continuous variable $y$ and a discrete one $n\in\mathbb Z$. The 1+1 reduction $\pal_y u_n=0$ of \eqref{tod1} gives the classical Toda lattice
$$
\ddot q_n =e^{q_{n+1}-q_n} -e^{q_n-q_{n-1}}, \quad u_n = q_{n+1}-q_n
$$
 i.e., an infinite system of points on the line with exponential interaction of the neighbors. The dispersionless limit is the PDE
\beq\label{tod2}
u_{tt} -u_{yy} = \left( e^u\right)_{xx}
\eeq
 for the function $u=u(x,y,t)$ obtained by interpolating
$$
u_n(y,t) = u(\epsilon\, n, y, t)
$$
rescaling
$$
y\mapsto \epsilon\, y, \quad t\mapsto \epsilon t
$$
and then setting $\epsilon\to 0$.

\subsection{\bf 2D Toda lattice as an integrable hierarchy with an infinite number of dependent variables}\par 
In order to develop an appropriate Frobenius manifolds technique we embed \eqref{tod1}, following K.Ueno and K.Takasaki \cite{ut}, into a hierarchy of an infinite number of flows associated with a pair of semi-infinite difference operators
\eqa\label{uta1}
&&
L=\qquad\quad\Delta + u_0(n) +u_{-1}(n) \Delta^{-1}+u_{-2}(n)\Delta^{-2}+\dots
\nn\\
&&
\\
&& 
\bar L = \bar u_{-1}(n) \Delta^{-1} + \bar u_0(n) + \bar u_1(n) \Delta +\bar u_2(n) \Delta^2+\dots .
\nn
\eeqa
Here $\Delta$ is the shift operator
$$
\Delta \, f_n =f_{n+1}.
$$
The hierarchy can be written in the familiar Lax form
\eqa\label{uta2}
&&
\frac{\pal L}{\pal s_k} =\left[ (L^k)_+, L\right], \quad \frac{\pal \bar L}{\pal s_k} =\left[ (L^k)_+, \bar L\right]
\nn\\
&&
\\
&&
\frac{\pal L}{\pal \bar s_k} =\left[ (\bar L^k)_-, L\right], \quad \frac{\pal \bar L}{\pal \bar s_k} =\left[ (\bar L^k)_-, \bar L\right]
\nn
\eeqa
$k\geq 1$. Here the positive/negative parts of a difference operator 
$$
M=\sum_{m\in \mathbb Z}  a_m \Delta^m
$$
are defined by
\eqa
&&
M_+ = \sum_{m=0}^\infty  a_m \Delta^m
\nn\\
&&
M_-=\sum_{m=-1}^{-\infty}  a_m \Delta^m.
\nn
\eeqa
In particular the coefficient
$$
u_n=\log \bar u_{-1}(n)
$$
as a function of $t=s_1-\bar s_1$, $y=s_1+\bar s_1$ satisfies \eqref{tod1}.

The hierarchy \eqref{uta2} possesses all usual properties of 1+1 systems: the flows \eqref{uta2} commute pairwise and admit a bihamiltonian description \cite{ca1}.
However, it contains an infinite number of unknown functions 
$$
u_i(n), ~ \bar u_j(n), \quad i\leq 0, ~ j\geq -1
$$ 
of the lattice variable $n$. Because of this one expects that an infinite dimensional Frobenius manifold must be brought under consideration.

Let us repeat that the Frobenius manifold structure is intrinsically involved in the description of the dispersionless limit of the hierarchy. The latter has been thoroughly studied in \cite{tt}. One has to replace the difference operators by their symbols; the coefficients $u_i$ and $\bar u_j$ become functions of the continuous spatial variable $x$ (cf. the above interpolation procedure); replacing in \eqref{uta2} the commutators of operators by the Poisson brackets of their symbols one obtains the flows of the dispersionless limit of the hierarchy (see eqs. \eqref{g-todalax} below). These equations have been extensively studied \cite{ber,its,kodama,mm,taka,teo,zabro} after the discovery, due to M.Mineev-Weinstein, P.B.Wiegmann and A.Zabrodin \cite{mwz,wz}, of a remarkable connection between the dispersionless 2D Toda hierarchy and the theory of conformal maps.

In our construction of the infinite-dimensional Frobenius manifold associated with the 2D Toda hierarchy we have to work with the same symbols of the {\it pair} of Lax operators $L$, $\bar L$. This is the main novelty with respect to the already widely accepted scheme of \cite{dvv} that suggests to use the symbol of the Lax operator as the ``Landau -- Ginzburg superpotential" in order to construct the Frobenius manifold (aka the small phase space of the two-dimensional topological field theory). For the 2D Toda case one has to deal with a pair of ``Landau -- Ginzburg superpotentials" treating them on equal footing.
The constructions heavily rely on the suitably chosen analytic properties of these symbols that we will now describe in order to proceed to the formulation of main results of the paper.

\medskip

\subsection{\bf Main results. A Frobenius manifold for the 2D Toda lattice}\par 

Let $S^1$ be the unit circle $|z|=1$ on the complex $z$-plane. Denote $D_0$ and $D_\infty$ resp. the inner and outer parts of $S^1$ on the Riemann sphere. Let ${\mathcal H}(D_0)$, resp. ${\mathcal H}(D_\infty)$ be the space of functions holomorphic on the closed disk $D_0$, resp. $D_\infty$, that is, functions holomorphic on $D_0$/$D_\infty$ admitting analytic continuation into bigger disks
$$
|z|<1+\rho\quad \mbox{or} \quad |z|>1-\rho\quad \mbox{respectively}
$$
for some positive $\rho$. 
Furthermore denote $\dot{\mathcal H}(D_0)$ and $\dot{\mathcal H}(D_\infty)$ extensions of these spaces allowing for a function to have simple poles at $z=0$ and $z=\infty$ resp. Functions in $\dot{\mathcal H}(D_\infty)$ will be denoted by
\beq\label{lam1}
\lambda(z) = u_1 z + u_0 + \frac{u_{-1}}{z}+\dots\in \dot{\mathcal H}(D_\infty),
\eeq
functions in $\dot{\mathcal H}(D_0)$ will be denoted\footnote{In this paper bar {\it never} stands for complex conjugation unless the opposite is explicitly stated. The coefficients of the Laurent expansions \eqref{lam1} and \eqref{lam2} can be considered as complex coordinates on the spaces $\dot{\mathcal H}(D_\infty)$ and $\dot{\mathcal H}(D_0)$ respectively.} by
\beq\label{lam2}
\bar\lambda(z) = \frac{\bar u_{-1}}{z} + \bar u_0 + \bar u_1 z +\dots \in \dot{\mathcal H}(D_0).
\eeq
Define an infinite-dimensional manifold $M$ as an affine subspace in the direct sum
\beq\label{mani}
M=\left\{ (\lambda, \bar\lambda)\in \dot{\mathcal H}(D_\infty)\oplus \dot{\mathcal H}(D_0)\, | \, u_1=1\right\}.
\eeq
The tangent space to $M$ at any point is isomorphic to the direct sum
\beq\label{tan}
TM=  {\mathcal H}(D_\infty) \oplus \dot{\mathcal H}(D_0)
\eeq
identifying first order linear differential operators with the derivatives of the functions $\lambda(z)$, $\bar\lambda(z)$
\beq\label{doper}
\pal \mapsto (\pal \lambda(z), \pal\bar\lambda(z))\in  {\mathcal H}(D_\infty) \oplus \dot{\mathcal H}(D_0) = TM
\eeq
(the differentiation with $z=\mbox{const}$). Similarly, the cotangent space is identified with
\beq\label{cotan}
T^*M =\dot{\mathcal H}(D_0) \oplus {\mathcal H}(D_\infty).
\eeq
It is understood that the duality between the tangent and cotangent spaces is established by the residue pairing 
\beq\label{pair}
\langle \hat\omega, \hat\alpha\rangle= \frac1{2\pi i} \oint_{|z|=1} \left[\alpha(z) \omega(z) + \bar\alpha(z)\bar\omega(z)\right]\, dz, \quad \hat\alpha=(\alpha,\bar\alpha)\in TM, \quad \hat\omega=(\omega, \bar\omega)\in T^*M.
\eeq
The following linear functionals will be useful in computations
\beq\label{dlam}
\langle d\lambda(p), \hat\alpha\rangle = \alpha(p), \quad \langle d\bar\lambda(p), \hat\alpha\rangle=\bar\alpha(p), \quad \hat\alpha=(\alpha,\bar\alpha)\in TM.
\eeq
Using Cauchy integral formula one obtains the following realization of these one-forms as elements of the space $T^*M= \dot{\mathcal H}(D_0) \oplus {\mathcal H}(D_\infty)$
\eqa\label{dlam1}
&&
d\lambda(p) =\left(\frac{p}{z} \frac1{p-z}, 0\right), \quad |z|< |p|
\nn\\
&&
\\
&&
d\bar\lambda(p) =\left(0, \frac{z}{p} \frac1{z-p}\right), \quad |z|>|p|.
\nn
\eeqa

We are now ready to define a Frobenius manifold structure on a suitable infinite dimensional submanifold $M_0$ of $M$ (see below). Recall that a Frobenius manifold must be equipped with a Frobenius algebra structure on the tangent bundle such that the associated nondegenerate symmetric invariant bilinear form $< ~,~>$ is a metric of vanishing curvature and the product of flat vector fields admits the representation
\beq\label{fro1}
<\pal_1\cdot \pal_2, \pal_3> =\pal_1 \pal_2\pal_3 F.
\eeq
Here $\pal_1$, $\pal_2$, $\pal_3$ are three arbitrary flat vector fields, the function $F$ is called the {\it potential} of the Frobenius manifold.
Besides the above conditions there must be a flat unit vector field and an Euler vector field involved in the quasihomogeneity condition (see details in \cite{icm}). 

Due to nondegeneracy of the invariant bilinear form, the induced isomorphism between tangent and cotangent bundles defines a Frobenius algebra structure also on the cotangent spaces. In our construction we begin just with the Frobenius algebra structure on $T^*M$.

Define a symmetric inner product and a multiplication on the cotangent space $T^*M$ at the point $(\lambda, \bar\lambda)$ by
\beq\label{eta3}
< d\alpha(p), d\beta(q)>_* =\frac{p\, q}{p-q} \left( \epsilon(\alpha)\, \beta'(q) -\epsilon(\beta)\, \alpha'(p)\right)
\eeq
and
\beq\label{dlam3}
d\alpha(p)\cdot d\beta(q) = \frac{p\, q}{p-q} \left[ \alpha'(p)\, d\beta(q) -\beta'(q)\, d\alpha(p)\right].
\eeq
Here $d\alpha(p)$, $d\beta(q)$ stand for one of the symbols $d\lambda(p)$ or $d\bar\lambda(p)$, the signs $\epsilon(\alpha)$, $\epsilon(\beta)$ are defined as follows
$$
\epsilon(\alpha)=1 \quad \mbox{if}\quad \alpha=\lambda\quad \mbox{and} \quad \epsilon(\alpha)=-1 \quad \mbox{if}\quad \alpha=\bar\lambda.
$$
As any 1-form $\hat\omega=(\omega(z), \bar\omega(z))$ can be represented as a linear combination of the 1-forms $d\lambda(p)$, $d\bar\lambda(p)$
$$
\hat\omega=\frac1{2\pi i} \oint_{|p|=1} \left(\omega(p) d\lambda(p)+  \bar\omega(p) d\bar\lambda(p)\right)dp,
$$
the inner product and the multiplication extend onto the entire cotangent bundle $T^*M$.

\begin{prop} \label{prop1} For any point $(\lambda,\bar\lambda)\in M$ the multiplication \eqref{dlam3} defines on $T^*_{(\lambda,\bar\lambda)}M$ a structure of a commutative associative algebra with an invariant bilinear form \eqref{eta3}. The latter does not degenerate on the open subset of $M$ defined by the conditions
\beq\label{pr1}
\bar u_{-1}\neq 0, \quad \lambda'(z) +\bar\lambda'(z) \neq 0 \quad {\rm for}\quad |z|=1.
\eeq
\end{prop}

We will now describe a submanifold $M_0\subset M$ on which the Frobenius structure will be introduced.
This will be an open subspace $M_0\subset M$ defined by the following conditions.
Denote
\beq\label{ww}
w(z) =\lambda(z) +\bar\lambda(z).
\eeq
First, for $(\lambda, \bar\lambda)\in M_0$ one must have $w'(z)\neq 0$ for any $z\in S^1$ and the image
$$
\Gamma=w(S^1)
$$
of the unit circle is required to be a non-selfintersecting positively oriented closed curve encircling the origin $w=0$. Second, we impose the condition $\bar u_{-1}\neq 0$. The manifold $M_0$ is fibered over the space $M_{\rm red}$ of parametrized simple analytic curves
\beq
M_0\ni (\lambda(z), \bar\lambda(z))  \mapsto \left\{ z\to w(z)\,|\, |z|=1\right\}\in M_{\rm red}
\eeq
with a two-dimensional fiber. One can choose 
\beq
u=\log \bar u_{-1}, \quad v=\bar u_0
\eeq
as coordinates on the fiber.

\begin{theorem}\label{t-main} The above formulae \eqref{eta3}, \eqref{dlam3} define on $M_0$ a structure of a semisimple infinite-dimensional Frobenius manifold with the unit vector 
\beq\label{unit1}
e=(-1,1)\in TM,
\eeq
the Euler vector field
\beq\label{euler}
E=\left( \lambda(z) -z\, \lambda'(z), \bar\lambda(z) -z\, \bar\lambda'(z)\right)\in T_{(\lambda, \bar\lambda)}M
\eeq
and the potential 
\eqa\label{pot}
&&
F=\frac12\frac1{(2\pi i)^2} {\oint\oint}_{|z_1|<|z_2|} \frac{w(z_1)}{z_1}\frac{w(z_2)}{z_2} \log\frac{z_2-z_1}{z_2}\, dz_1\, dz_2
\nn\\
&&
\\
&&
+\frac12 (\bar u_0 - u_0) \left[ \frac1{2\pi i} \oint_{|z|=1} \frac{w(z)}{z} \log\frac{w(z)}{z}\, dz - u_0 - \bar u_0\right]
\nn\\
&&
+\frac12 \bar u_0^2 \log \bar u_{-1} +\bar u_{-1} +u_{-1} +\bar u_{-1} \bar u_1.
\nn
\eeqa
\end{theorem}

Let us describe the canonical coordinates \cite{icm} on the semisimple
part of the Frobenius manifold $M_0$. Consider the analytic curve 
\beq\label{sig1}
\Sigma:= \{ S^1\ni p\mapsto \frac{\lambda'(p)}{\lambda'(p) +\bar\lambda'(p)}\}.
\eeq
Denote $M^0_{s\, s}\subset M_0$ the subset consisting of pairs $(\lambda, \bar\lambda)$ such that the curve $\Sigma$ is smooth non-self intersecting. For a given curve $\Sigma$ introduce the following functional on $M_{s\, s}^0$ depending on the point of the curve
\beq\label{sig2}
u_\sigma:= \left[\sigma\,\bar\lambda(p) +(\sigma-1)\, \lambda(p)\right]_{p=p(\sigma)}, \quad \sigma\in\Sigma
\eeq
where $p=p(\sigma)\in S^1$ is determined from the equation
\beq\label{sig3}
\left[\sigma\,\bar\lambda'(p) +(\sigma-1)\, \lambda'(p)\right]_{p=p(\sigma)}=0,\quad \sigma\in\Sigma.
\eeq

\begin{prop} \label{prop4}The functionals $u_\sigma$ are canonical coordinates on $M_{s\, s}^0$.
\end{prop}

\begin{remark} For the dispersionless limit of the 1+1 Lax equations the well known prescription suggests to take the critical values of the symbol of the Lax operator in order to obtain the Riemann invariants (aka the canonical coordinates) of the dispersionless equations. This rule extends also to the 1+1 Whitham equations, where the Riemann invariants are given by the ramification points of the spectral curve \cite{ffm}. Our procedure \eqref{sig2}, \eqref{sig3} looks very similarly. The main difference is that now the Riemann invariants are labeled by a continuous parameter running through the curve $\Sigma$.
\end{remark}

The proofs of Propositions \ref{prop1}, \ref{prop4} and Theorem \ref{t-main} will be given in Section \ref{sec2}. In particular, an expression for the metric induced by \eqref{eta3} on the tangent bundle is given in the formula \eqref{eta4} below, the flat coordinates for the metric can be found in \eqref{rh} - \eqref{flat}, the symmetric trilinear form $<\pal_1\cdot \pal_2, \pal_3>$ on $TM_0$ (i.e., the would-be triple correlator  \cite{wi} of the primary fields of the associated 2D topological field theory with an infinite number of primaries) is written in \eqref{corr}.

Recall \cite{icm} that on the cotangent bundle of an arbitrary Frobenius manifold there exists another important symmetric bilinear form defined by the formula
\beq\label{inter}
(\omega_1, \omega_2)_* = {\rm i}_E(\omega_1\cdot \omega_2)
\eeq
(the so-called {\it intersection form} of the Frobenius manifold). It does not degenerate outside a closed analytic subset; the curvature of the induced metric vanishes. For the case under consideration the intersection form admits the following explicit expression.

\begin{prop}\label{prop2} The intersection form of the Frobenius manifold $M_0$ is given by the formula
\beq\label{inter1}
\left(d\alpha(p), d\beta(q)\right)_* = \frac{p\, q}{p-q} \left[ \alpha'(p)\beta(q) -\beta'(q)\alpha(p)\right] +p\, q\, \alpha'(p) \beta'(q).
\eeq
It does not degenerate on the open subset of $M_0$ defined by the conditions
\beq\label{inter4}
\lambda'(z)\neq0, \quad \bar\lambda'(z)\neq 0, \quad \lambda(z)\bar\lambda'(z)-\bar\lambda(z) \lambda'(z)\neq 0 \quad {\rm for}\quad |z|=1.
\eeq
On this subset it defines a metric of zero curvature given by the following inner product on the tangent space
\beq\label{inter5}
(\pal_1, \pal_2) =\frac1{2\pi i} \oint_{|z|=1} \frac{\left( \frac{\pal_1\lambda}{\lambda'} - \frac{\pal_1\bar\lambda}{\bar\lambda'}\right)\left( \frac{\pal_2\lambda}{\lambda'} - \frac{\pal_2\bar\lambda}{\bar\lambda'}\right)}{\frac{\lambda}{\lambda'} -\frac{\bar\lambda}{\bar\lambda'}}\, \frac{dz}{z^2}.
\eeq
\end{prop}

The notations in the formula \eqref{inter1} are similar to those in \eqref{eta3}. The complement  locus to the subset \eqref{inter4} is the discriminant of the infinite-dimensional Frobenius manifold.

\subsection{\bf The infinite-dimensional Frobenius manifold and the extended dispersionless 2D Toda hierarchy}\par

The flat pencil of metrics \eqref{eta3} and \eqref{inter1} plays an important role in the bihamiltonian formulation of the associated integrable hierarchy. Recall that the dispersionless integrable hierarchy  associated with a given $n$-dimensional Frobenius manifold $M$ (the so-called {\it Principal Hierarchy}) is an infinite family of pairwise commuting flows on the (formal) loop space
$$
\cL M: =\left\{ S^1 \to M\right\}.
$$
All the equations of the hierarchy are evolutionary PDEs with one spatial and one time variable. In the flat coordinates $v^1$, \dots, $v^n$ the equations of the lowest level of the hierarchy (the so-called {\it primary flows}) can be written in the form
\beq\label{inter2}
\frac{\pal v^\gamma}{\pal t^{\alpha,0}} =\sum_{\beta=1}^n c_{\alpha\beta}^\gamma(v)\, \frac{\pal v^\beta}{\pal x}, \quad \gamma=1, \dots, n
\eeq
(the number $\alpha$ of this equation may take values from $1$ to $n$). Here $c_{\alpha\beta}^\gamma(v)$ are the structure constants of the Frobenius algebra on $TM$. The equations $\pal v^\gamma/ \pal t^{\alpha, p} =\dots$ of the higher levels $p>0$ are obtained by a suitable recursion procedure (see details in \cite{normal,icm}).

Thus, an infinite number of primary flows must be constructed for an infinite-dimensional Frobenius manifold. Only two of these flows are covered by the dispersionless limit of the 2D Toda equations \eqref{uta2}, namely,
$$
\frac{\pal}{\pal s_1} =-\frac{\pal}{\pal t^{0,0}}+\frac{\pal}{\pal t^{u,0}}, \quad \frac{\pal}{\pal \bar s_1} =-\frac{\pal}{\pal t^{u,0}}.
$$
All other primary flows have to be constructed. They are described in the following

\begin{theorem}\label{t-main2}
The primary flows of the Principal Hierarchy associated with the infinite-dimensional Frobenius manifold of Theorem \ref{t-main} have the following form
\eqa\label{inter3}
&&
\frac{\pal \lambda(z)}{\pal t^{\alpha,0}} =\frac1{\alpha+1} \left\{ \left(w^{\alpha+1}(z)\right)_{<0}, \lambda(z) \right\}, \quad \frac{\pal \bar\lambda(z)}{\pal t^{\alpha,0}} =-\frac1{\alpha+1} \left\{ \left(w^{\alpha+1}(z)\right)_{\geq 0}, \bar\lambda(z) \right\}
\nn\\
&&
 \alpha \in \mathbb Z, \quad \alpha\neq -1,
\nn\\
&&
\\
 &&
\frac{\pal \lambda(z)}{\pal t^{-1,0}}= \left\{\left( \log\frac{w(z)}{z}\right)_{<0}+ \log z ,\lambda(z)\right\}, \quad \frac{\pal \bar\lambda(z)}{\pal t^{-1,0}}=-\left\{ \left( \log\frac{w(z)}{z}\right)_{\geq 0}, \bar\lambda(z)\right\}
\nn\\
&&
\frac{\pal }{\pal t^{v,0}}=\frac{\pal }{\pal x}
\nn\\
&&
\frac{\pal }{\pal t^{u,0}}=-\frac{\pal}{\pal \bar s_1}
\nn
\eeqa
All these flows are symmetries of the dispersionless limit \eqref{g-todalax} of the 2D Toda hierarchy.
\end{theorem} 

In these formulae the dispersionless Lax representation of the primary flows is given, the curly bracket stands for the standard Poisson bracket on the cylinder $(z,x)\in S^1 \times \mathbb R$:
$$
\left\{ f(z,x), g(z,x)\right\} =  z \frac{\pal f}{\pal z} \frac{\pal g}{\pal x} - z \frac{\pal g}{\pal z} \frac{\pal f}{\pal x}.
$$

The proofs of the Proposition \ref{prop2} and the Theorem \ref{t-main2} are given in Section \ref{sec3} where  we will also add more explanations about connections between our infinite-dimensional Frobenius manifold and the 2D Toda lattice. We also construct an analogue of the Riemann invariants for the primary flows \eqref{inter3}:

\begin{prop} \label{prop3} The primary flows \eqref{inter3} in the canonical coordinates \eqref{sig2} take the following diagonal form
\eqa\label{riem}
&&
\frac{\pal u_\sigma}{\pal t^{i,0}} =A_i(\sigma)\, \frac{\pal u_\sigma}{\pal x},\quad i\in \mathbb Z, \quad \sigma\in \Sigma
\nn\\
&&
\\
&&
A_i(\sigma) = -p(\sigma)\left[ \sigma\, (w^i(p) w'(p))_{\geq 0} +(\sigma-1)\, (w^i(p) w'(p))_{\leq -1}\right]_{p=p(\sigma)}
\nn\\
&&
\frac{\pal u_\sigma}{\pal t^{u,0}} =A_u(\sigma)\, \frac{\pal u_\sigma}{\pal x}, \quad A_u(\sigma) =\frac{e^u}{p(\sigma)}.
\nn
\eeqa
\end{prop}

The explicit realization of higher flows of the Principal Hierarchy associated with the Frobenius manifold $M_0$ will be given in a separate publication.

\begin{remark} In \cite{bmrwz,krich}  it was shown that the logarithm of the tau-function $\tau ({\bf s}, {\bf \bar s})$ of any solution to the dispersionless 2D Toda hierarchy satisfies the WDVV equation. This gives solutions to WDVV depending on an infinite number of variables. A particular tau-function admits an elegant realization on the space of simply connected plane domains bounded by simple analytic contours assuming the possibility to locally parametrize the domains by their exterior harmonic moments (see also \cite{takh}). Such an assumption has been rigorously justified in \cite{ef} for the class of polynomial boundary curves in which case all harmonic moments but a finite number are equal to zero. A connection between our solution to WDVV given in \eqref{pot} and the one of \cite{bmrwz,krich} has to be clarified yet. It is clear however that the two WDVV solutions are defined on different spaces.
\end{remark}

\begin{remark} In \cite{am} M.Adler and P. van Moerbeke proposed an extension of the 2D Toda hierarchy by adding the flows with the Lax representation of the form
\beq\label{avm1}
\frac{\pal L}{\pal s_{ij}} = \left[ P_+ , L\right], \quad \frac{\pal \bar L}{\pal s_{ij}}
=-\left[ P_-, \bar L\right]
\eeq
with
\beq\label{avm2}
P=L^i \bar L^j, \quad i, \, j\geq 0.
\eeq
They argued that these flows, if well-defined, should commute pairwise. Note that the dispersionless limits of these flows make sense on our infinite-dimensional Frobenius manifold since the products $\lambda^i(z) \bar\lambda^j(z)$ are well defined for all nonnegative integers $i$, $j$. One can check that all these dispersionless flows are linear combinations of the flows of the Principal Hierarchy associated with the Frobenius manifold $M_0$.
\end{remark}

%

\section{Construction of a Frobenius manifold}\label{sec2}\par

Let us begin with the proof of Proposition \ref{prop1}.
Define a multiplication on the cotangent space $T^*_{(\lambda, \bar\lambda)}M$ by the following formula
\eqa\label{prod}
&&
\hat\omega_1\cdot \hat\omega_2=
\\
&&
=z^2 \left( \omega_1 (\lambda' \omega_2 +\bar\lambda'\bar\omega_2)_{\geq -1} +\omega_2 (\lambda' \omega_1 +\bar\lambda'\bar\omega_1)_{\geq -1}-[\lambda'\omega_1\omega_2 +\bar\lambda'(\omega_1\bar\omega_2+\bar\omega_1\omega_2)]_{\geq -3},
\right.
\nn\\
&&
\nn\\
&&
\left.
-\bar\omega_1 (\lambda'\omega_2+\bar\lambda'\bar\omega_2)_{\leq -2} - \bar\omega_2 (\lambda'\omega_1+\bar\lambda'\bar\omega_1)_{\leq -2} +[\bar\lambda'\bar\omega_1 \bar\omega_2 +\lambda'(\omega_1\bar\omega_2 +\bar\omega_1\omega_2)]_{\leq -2}
 \right).
 \nn
\eeqa
Here we use the notations $\left( f\right)_{\geq k} $ and $\left( f\right)_{\leq k} $ for the projections of a function 
$$
f=f(z)=\sum_{n\in \mathbb Z} f_n z^n
$$ 
analytic on a neighborhood of $S^1$ defined by
\eqa\label{proj1}
&&
\left( f\right)_{\geq k}=\sum_{n\geq k} f_n z^n=\quad\frac{z^k}{2\pi i} \oint_{|z|<|\zeta|} \frac{\zeta^{-k} f(\zeta)}{\zeta-z} d\zeta
\\
&&
\left( f\right)_{\leq k}=\sum_{n\leq k} f_n z^n=-\frac{z^{k+1}}{2\pi i} \oint_{|z|>|\zeta|} \frac{\zeta^{-k-1} f(\zeta)}{\zeta-z} d\zeta
\label{proj2}
\eeqa
for a given integer $k$. Observe the simple identity
\beq\label{ide}
\oint dz\, f(z)\, \left(g(z)\right)_{\geq k} = \oint dz\, \left(f(z)\right)_{\leq -k-1}  g(z)
\eeq
often used in subsequent calculations. 

\begin{lemma}\label{lem20} The product of the linear functionals of the form \eqref{dlam} with respect to the multiplication \eqref{prod} coincides with \eqref{dlam3}.
\end{lemma}

\pf Let us compute the product of 1-forms
$$
\hat\omega_1=d\lambda(p), \quad \hat\omega_2=d\lambda(q).
$$
In this case the product $\hat\omega_1\cdot \hat\omega_2\in \dot{\cal H}(D_0)$. We have
$$
\left( \lambda' \omega_2\right)_{\geq -1} =\frac1{2\pi i z} \oint_{|z|<|\zeta|<|q|} \frac{q\,\lambda'(\zeta)}{(q-\zeta) (\zeta-z)} d\zeta =-\frac{q}{z} \left[ \res_{\zeta=q} +\res_{\zeta=\infty}\right] \frac{\lambda'(\zeta)}{(q-\zeta) (\zeta-z)} d\zeta =\frac{q}{z} \frac{\lambda'(q)}{q-z},
$$
\eqa
&&
\left( \lambda'\omega_1 \omega_2\right)_{\geq -3} =\frac1{2\pi i z^3} \oint_{|z|<|\zeta|<\min (|p|, |q|)} \frac{p\, q\, \zeta\, \lambda'(\zeta)}{(p-\zeta)(q-\zeta)(\zeta-z)} d\zeta
\nn\\
&&
=-\frac{pq}{z^3}\left[ \res_{\zeta=p}+\res_{\zeta=q} +\res_{\zeta=\infty}\right]\frac{\zeta\, \lambda'(\zeta)}{(p-\zeta)(q-\zeta)(\zeta-z)} d\zeta
=\frac{pq}{z^3}\left[ \frac{p\, \lambda'(p)}{(q-p)(p-z)} +\frac{q\, \lambda'(q)}{(p-q)(q-z)}\right].
\nn
\eeqa
So
\eqa
&&
\hat\omega_1\cdot \hat\omega_2 =z^2 \left( \omega_1 \left(\lambda'\omega_2\right)_{\geq -1} +  \omega_2 \left(\lambda'\omega_1\right)_{\geq -1} -\left( \lambda' \omega_1 \omega_2\right)_{\geq -3}, 0\right) 
\nn\\
&&
=
pq \left( \frac{\lambda'(q)}{(p-z)(q-z)} +\frac{\lambda'(p)}{(p-z(q-z)} -\frac1{z}\left[ \frac{p\,\lambda'(p)}{(q-p)(p-z)} +\frac{q\, \lambda'(q)}{(p-q)(q-z)}\right], 0\right)
\nn\\
&&
=\left( \frac{p\, q}{p-q} \left[ \lambda'(p) \, \frac{q}{z} \frac1{q-z} -\lambda'(p)\, \frac{p}{z} \frac1{p-z}\right], 0\right)= \frac{p\, q}{p-q}\left[ \lambda'(p)\, d\lambda(q) -\lambda'(q)\, d\lambda(p)\right].
\nn
\eeqa
Next, for the product of $\hat\omega_1=d\lambda(p)$, $\hat\omega_2=d\bar\lambda(q)$ we have the following expression
$$
\hat\omega_1\cdot\hat\omega_2=z^2\left( \omega_1 \left(\bar\lambda'\bar\omega_2\right)_{\geq -1} -\left( \bar\lambda'\omega_1 \bar\omega_2\right)_{\geq -3}, -\bar\omega_2 \left( \lambda' \omega_1\right)_{\leq -2} +\left( \lambda'\omega_1 \bar\omega_2\right)_{\leq -2}\right).
$$
The computation similar to the one above gives
$$
\left( \bar\lambda'\bar\omega_2\right)_{\geq -1} =\frac1{2\pi i z\, q} \oint_{\max(|q|, |z|)<|\zeta|}  \frac{\zeta^2\bar\lambda'(\zeta)}{(\zeta-q)(\zeta-z)} d\zeta =\frac{q^2 \bar\lambda'(q) - z^2 \bar\lambda'(z)}{q\, z (q-z)},
$$
\eqa
&&
\left( \bar\lambda'\omega_1 \bar\omega_2\right)_{\geq -3} =\frac1{2\pi i z} \frac{p}{q} \oint_{\max (|q|, |z|)<|\zeta|<|p|} \frac{\zeta^2\bar\lambda'(\zeta)}{(p-\zeta)(\zeta-q)((\zeta-z)} d\zeta
\nn\\
&&
=\frac1{z^3} \frac{p}{q} \left[ \res_{\zeta=q} + \res_{\zeta=z} +\res_{\zeta=0}\right] \frac{\zeta^2\bar\lambda'(\zeta)}{(p-\zeta)(\zeta-q)((\zeta-z)} d\zeta = \frac1{z^3} \frac{p}{q} \left[ \frac{q^3 \bar\lambda'(q)}{(p-q)(q-z)} + \frac{z^3 \bar\lambda'(z)}{(p-z)(z-q)}\right].
\nn
\eeqa
Putting these two terms together and multiplying by $z^2$ we arrive at the expression for the first component of the product
$$
=-\frac{p\, q}{p-q} \bar\lambda'(q) \frac{p}{z}\frac1{p-z}= -\frac{p\, q}{p-q} \bar\lambda'(q)\, d\lambda(p).
$$
For the second component the computation is similar
\eqa
&&
\left( \lambda' \omega_1\right)_{\leq -2} =-\frac{p}{2\pi i z} \oint_{|\zeta| < \min(|z|, |p|)} \frac{\lambda'(\zeta)}{(p-\zeta)(\zeta-z)} d\zeta =\frac{p}{z} \left[\res_{\zeta=p}+\res_{\zeta=z}+\res_{\zeta=\infty}\right] \frac{\lambda'(\zeta)}{(p-\zeta)(\zeta-z)} d\zeta
\nn\\
&&
 \hskip 1.7truecm =\frac{p}{z} \frac{\lambda'(z)-\lambda'(p)}{p-z},
\nn
\eeqa
\eqa
&&
\left( \lambda' \omega_1 \bar\omega_2\right)_{\leq -2} =-\frac1{2\pi i z} \frac{p}{q} \oint_{|q| <|\zeta| < \min(|p|, |z|)} \frac{\zeta\, \lambda'(\zeta)}{(p-\zeta)(\zeta-q)(\zeta-z)} d\zeta
\nn\\
&&
 =\frac{p}{q}\frac1{z} \left[ \res_{\zeta=p} +\res_{\zeta=z}+\res_{\zeta=\infty}\right] \frac{\zeta\, \lambda'(\zeta)}{(p-\zeta)(\zeta-q)(\zeta-z)} d\zeta =\frac{p}{q}\frac1{z(p-z)} \left[\frac{z\, \lambda'(z)}{z-q}-\frac{p\, \lambda'(p)}{p-q}\right].
\nn
\eeqa
Putting these together and multiplying by $z^2$ we arrive at
$$
z^2 \left( -\bar\omega_2 \left( \lambda' \omega_1\right)_{\leq -2} +\left( \lambda'\omega_1 \bar\omega_2\right)_{\leq -2}\right)=\frac{p\, q}{p-q} \, \lambda'(p) \frac{z}{q} \frac1{z-q} =\frac{p\, q}{p-q} \, \lambda'(p) \, d\bar\lambda(q).
$$
We have proved that
$$
d\lambda(p)\cdot d\bar\lambda(q)= \frac{p\, q}{p-q}\left[ \lambda'(p) \, d\bar\lambda(q) -\bar\lambda'(q)\, d\lambda(p)\right].
$$
The computation of the product $d\bar\lambda(p)\cdot d\bar\lambda(q)$ is similar. We leave it as an exercise for the reader. \epf

\begin{lemma} \label{lm21} For any $(\lambda, \bar \lambda)\in M$ the formula \eqref{prod} defines on $T^*_{(\lambda, \bar\lambda)}M$ a structure of a commutative associative algebra.
\end{lemma}

\pf Commutativity of the product \eqref{prod} is obvious. In order to prove associativity let us compute the product of three 1-forms of the form \eqref{dlam}. An easy computation shows that this product  can be written in the following manifestly symmetric way
\eqa\label{produ3}
&&
[d\alpha(p)\cdot d\beta(q)]\cdot d\gamma(r) = 
\\
&&=\frac{p\, r}{p-r} \frac{p\, q}{p-q} \beta'(q)\gamma'(r)\, d\alpha(p)
+\frac{q\, p}{q-p} \frac{q\, r}{q-r} \alpha'(p)\gamma'(r) \, d\beta(q) 
\nn\\
&&
 +\frac{r\, p}{r-p} \frac{r\, q}{r-q} \alpha'(p) \beta'(q)\, d\gamma(r).
\nn
\eeqa
\epf

\begin{remark} On an open subset of $M$ defined by the condition
$$
\bar u_{-1}\neq 0
$$
the algebra \eqref{prod} will have
a unit
\beq\label{unit}
e^*=\left(0,\frac1{\bar u_{-1}}\right). 
\eeq
\end{remark}

\begin{remark} The following formula generalizing \eqref{produ3} can be easily derived by induction:
\beq\label{produi}
d\alpha_1(p_1)\cdot \dots \cdot d\alpha_n (p_n) =\sum_{i=1}^n \frac{\alpha_1'(p_1)}{p_i^{-1}-p_1^{-1}} \frac{\alpha_2'(p_2)}{p_i^{-1}-p_2^{-1}}\dots d\alpha_i(p_i) \dots \frac{\alpha'_n(p_n)}{p_i^{-1} -p_n^{-1}}.
\eeq
\end{remark}

Introduce a linear map
$$
\eta : T^*_{(\lambda, \bar\lambda)}M \to T_{(\lambda, \bar\lambda)}M
$$
by the formula
\eqa\label{eta1}
&&
\eta(\hat\omega) = z^2\left(  ( \lambda'\omega+\bar\lambda' \bar\omega)_{\leq -2} -\lambda' (\omega -\bar\omega)_{\leq -2} ,  ( \lambda'\omega+\bar\lambda' \bar\omega)_{\geq -1} +\bar\lambda'  (\omega -\bar\omega)_{\geq -1}\right)
\nn\\
&&
\\
&&
 \quad \hat\omega=(\omega,\bar\omega)\in T_{(\lambda,\bar\lambda)}^*M.
 \nn
\eeqa

\begin{lemma} The associated bilinear form on $T^*_{(\lambda, \bar\lambda)}M$
\beq\label{eta2}
< \hat\omega_1, \hat\omega_2 >_* = \langle \hat\omega_1 , \eta( \hat\omega_2
) \rangle, \qquad \hat\omega_1, \hat\omega_2 \in T^*_{(\lambda,\bar\lambda)} M
\eeq
coincides with \eqref{eta3}.
\end{lemma}

\pf is obtained by a simple computation, similar to the one in the
proof of Lemma 2.1. \epf

\begin{lemma} The bilinear form \eqref{eta3} is invariant with respect to the multiplication \eqref{dlam3}:
\beq\label{inv}
< \hat\omega_1 \cdot \hat\omega_2, \hat\omega_3>_* = 
< \hat\omega_1,   \hat\omega_2\cdot \hat\omega_3>_*.
\eeq
for any $\hat\omega_1, \, \hat\omega_2, \, \hat\omega_3\in T^*_{(\lambda, \bar\lambda)}M$.
\end{lemma}

\pf As in the proof of Lemma \ref{lm21} let us compute the inner product \eqref{inv} choosing the three 1-forms $\hat\omega_1$, $\hat\omega_2$, $\hat\omega_3$ among $d\lambda(p)$ and $d\bar\lambda(p)$. Using the formula \eqref{eta1} one easily obtains the following symmetric expression
\eqa
&&
< d\alpha(p)\cdot d\beta(q), d\gamma(r)>_* =
\nn\\
&&
=-p\, q\, r\left[ \frac{\epsilon(\alpha) \beta'(q)\gamma'(r)p}{(p-q)(p-r)} +\frac{\epsilon(\beta)\alpha'(p)\gamma'(r)q}{(q-p)(q-r)} +\frac{\epsilon(\gamma) \alpha'(p)\beta'(q) r}{(r-p)(r-q)}\right].
\nn
\eeqa
\epf
Let us now prove nondegeneracy of the symmetric bilinear form \eqref{eta3} on the subspace defined in \eqref{pr1}. Recall that on this subspace one has $w'(z) \equiv \lambda'(z) +\bar \lambda'(z)\neq 0$ for $z\in S^1$. Moreover, the coefficient $\bar u_{-1}$ does not vanish.

\begin{lemma}\label{lm27} For any $(\lambda, \bar\lambda)$ satisfying \eqref{pr1} the linear operator \eqref{eta1} is an isomorphism.
\end{lemma}

\pf
We need to solve the equation
$$
\eta(\hat \omega)=\hat\alpha, \quad \hat\alpha=(\alpha, \bar\alpha)\in
T_{(\lambda, \bar\lambda)}M ,
$$
which is equivalent to the following system
\eqa
&& \alpha = z^2 ( \lambda' \omega + \bar\lambda' \bar\omega )_{\leq -2}  +
z^2 \lambda' (\bar\omega)_{\leq -2}, \nn \\
&& \bar\alpha = z^2 ( \lambda' \omega + \bar\lambda' \bar\omega )_{\geq -1}
+ z^2 \bar\lambda' \omega - z^2 \bar\lambda' (\bar\omega)_{\geq -1}, \nn
\eeqa
where we have used the assumptions $\alpha, \bar\omega \in
\mathcal{H}(D_{\infty})$ and $\bar\alpha, \omega \in \dot\mathcal{H}(D_0)$.
It easily follows that
\beq
\alpha + \bar\alpha = z^2 w' (\omega + \bar\omega_{\leq -2} ) \nn .
\eeq
Dividing by $z^2 w'$ and taking suitable projections one obtains
\eqa\label{inver}
&&
\omega=\frac1{z^2}\left(\frac{\alpha(z)+\bar\alpha(z)}{w'(z)}\right)_{\geq
1}
\nn\\
&&
\\
&&
\bar\omega=\frac1{z^2}\left(\frac{\alpha(z)+\bar\alpha(z)}{w'(z)}\right)_{\leq
2}+\frac1{\bar u_{-1}} \left( \frac{\bar\alpha_{-1}}{z}
+\bar\alpha_0\right).
\nn
\eeqa \epf

\begin{corollary} For $(\lambda, \bar\lambda)\in M$ satisfying \eqref{pr1} the non-degenerate symmetric bilinear form \eqref{eta3} on $T^*_{(\lambda, \bar\lambda)}M$ induces a non-degenerate symmetric bilinear form on $T_{(\lambda, \bar\lambda)}M$.  The latter can be written in the following form
\beq\label{eta4}
< \pal_1, \pal_2> =\frac1{2\pi i} \oint_{|z|=1} \frac{\pal_1 w(z) \, \pal_2 w(z)}{z^2w'(z)} \, {dz} - \res_{z=0} \frac{\pal_1 \ell(z) \, \pal_2 \ell(z)}{z^2\ell'(z)}{dz}
\eeq
for any two tangent vectors $\pal_1, \, \pal_2 \in  T_{(\lambda, \bar\lambda)}M$
where
\beq\label{ell}
\ell(z) =z+v+\frac{e^u}{z}.
\eeq
\end{corollary}

From the above lemmata it immediately follows the validity of the Proposition \ref{prop1}.

\medskip

Let us now proceed to the proof of Theorem \ref{t-main}.
Recall that the subspace $M_0\subset M$ is defined by the following conditions. 
First, for $(\lambda, \bar\lambda)\in M_0$ one must have $w'(z)\neq 0$ for any $z\in S^1$. Moreover, the image
$$
\Gamma=w(S^1)
$$
of the unit circle is required to be a non-selfintersecting positively oriented closed curve. Second, we impose the condition $\bar u_{-1}\neq 0$. It will also be technically  convenient to assume that the curve $\Gamma$ encircles the origin $w=0$.

The function $w(z)=\lambda(z)+\bar\lambda(z)$ is holomorphic on some neighborhood of the unit circle $S^1$. The functions $\lambda(z)$, $\bar\lambda(z)$ can be reconstructed from the triple $w(z), \bar u_0, \bar u_{-1}$ by the following formulae
\eqa\label{dubl1}
&&
\lambda(z) =\left( w(z)\right)_{\leq 0} +z -\bar u_0 - \frac{\bar u_{-1}}{z}
\nn\\
&&
\\
&&
\bar\lambda(z) =\left( w(z)\right)_{\geq 1} -z +\bar u_0 + \frac{\bar u_{-1}}{z}.
\nn
\eeqa
So, the coefficients $w_n$ of the Laurent expansion
$$
w(z) =\sum_{n\in \mathbb Z} w_n z^n
$$
along with $\bar u_0$ and $\bar u_{-1}$ can be used as an alternative system of coordinates on the manifold $M$.
The symmetric bilinear form on tangent spaces to the manifold $M_0$ will be called {\it metric} on this manifold.
We begin with proving that this metric is flat. To this end we will introduce a system of {\it flat coordinates} for this metric. They are obtained from the following procedure. Consider the inverse function
$$
z=z(w): \Gamma\to S^1.
$$
It is holomorphic on some neighborhood of the curve $\Gamma$ and satisfies
$$
|z(w)|_{w\in \Gamma}=1.
$$
Introduce the Riemann--Hilbert factorization of this function
\beq\label{rh}
z(w) = f_0^{-1}(w) f_\infty(w)\quad \mbox{for}\quad w\in \Gamma
\eeq
where the functions $f_0(w)$ and $f_\infty(w)/w$ are holomorphic and non-vanishing inside/outside the curve $\Gamma$ (in both cases holomorphicity can be assumed in a bigger domain containing the curve itself). The factorization will be uniquely defined by normalizing
$$
f_\infty(w) = w +O(1), \quad |w|\to\infty.
$$
Denote $t_n$ the coefficients of the Taylor expansions of the logarithms of these functions
\eqa\label{flat}
&&
\log f_0(w) = -t_0 - t_1 w - t_2 w^2 -\dots, \quad |w| \to 0
\nn\\
&&
\\
&&
\log\frac{f_\infty(w)}{w}=\frac{t_{-1}}{w} +\frac{t_{-2}}{w^2}+\dots, \quad |w|\to\infty.
\nn
\eeqa
These coefficients along with
\beq\label{flat1}
v=\bar u_0, \quad u=\log \bar u_{-1}
\eeq
constitute one more system of coordinates $({\bf t}, u, v)$ on $M_0$. In order to compute the curve in $M_0$ obtained by variation of the coordinate $t_k\mapsto t_k +\Delta t_k$ one has to invert the function
$z(w) e^{\Delta t_k\, w^k}$ and then restrict it to the unit circle. This implies the obvious formula
\beq\label{wc}
\frac{\pal}{\pal t_n} z(w)=w^n z(w)
\eeq
and hence
\beq\label{zc}
\frac{\pal}{\pal t_n} w(z) =-z\,w^n(z) w'(z)
\eeq
(note that in the formula \eqref{wc} we differentiate keeping $w={\rm const}$ while in \eqref{zc}  $z={\rm const}$). Therefore
\eqa\label{lamt}
&&
\frac{\pal\lambda(z)}{\pal t_n} =-z\,\left[w^n(z) w'(z)\right]_{\leq -1}
\nn\\
&&
\\
&&
\frac{\pal\bar\lambda(z)}{\pal t_n} =-z\left[w^n(z) w'(z)\right]_{\geq 0}.
\nn
\eeqa
Moreover,
\eqa\label{lamuv}
&&
\frac{\pal \lambda(z)}{\pal v}=-1, \quad \frac{\pal \bar\lambda(z)}{\pal v}=1
\nn\\
&&
\\
&&
\frac{\pal \lambda(z)}{\pal u}=-\frac{e^u}{z}, \quad \frac{\pal \bar\lambda(z)}{\pal u}=\frac{e^u}{z}.
\nn
\eeqa

\begin{lemma} The Gram matrix of the metric \eqref{eta2} becomes constant in the coordinates $({\bf t}, u, v)$, namely
\beq\label{flat5}
\left\langle \frac{\pal}{\pal t_k}, \frac{\pal}{\pal t_l}\right\rangle =\delta_{k+l, -1}, \quad \left<\frac{\pal}{\pal u}, \frac{\pal}{\pal v}\right> =1,
\eeq
all other inner products vanish.
\end{lemma}

\pf Let us use the formula \eqref{eta4}. With the help of \eqref{zc} we obtain
$$
\left\langle\frac{\pal}{\pal t_k}, \frac{\pal}{\pal t_l}\right\rangle=\frac1{2\pi i} \oint w^{k+l} w' dz=
\frac1{2\pi i} \oint w^{k+l} dw=\delta_{k+l,-1}.
$$
The second part of the formula is proved in a similar way. \epf

\begin{remark} Note the identity
\beq\label{kvad}
\frac1{2\pi i}\oint \frac{w(z)}{z} \log  \frac{w(z)}{z}\, dz = \frac12 \sum_{i+j=-1} t_i t_j -t_{-1}
\eeq
from which the convergence of the infinite sum in the right hand side follows. Another useful formula:
\beq\label{kvad1}
u_0= -t_{-1}-v.
\eeq
\end{remark}

\begin{lemma} The function \eqref{pot} on $M_0$ coincides with
\eqa\label{pot1}
&&
F({\bf t}, u, v) =\frac12 \frac1{(2\pi i)^2}\oint_\Gamma\oint_\Gamma {\rm Li}_3 \left( \frac{z(w_1)}{z(w_2)}\right)\, dw_1\, dw_2 +\frac1{2\pi i} \oint_\Gamma \left( \frac{e^u}{z(w)} - z(w)\right)\, dw
\nn\\
&&
\\
&&
\quad\quad\qquad +\frac1{4\pi i} \left( v +\frac12 t_{-1}\right) \oint_\Gamma \left(\log\frac{z(w)}{w}\right)^2 \, dw 
+\frac12 v^2 u -e^u.
\nn
\eeqa
\end{lemma}

Recall that the tri-logarithm is defined by its Taylor expansion
$$
{\rm Li}_3(x) =\sum_{k\geq 1} \frac{x^k}{k^3}, \quad |x|<1.
$$
The double integral must be regularized in such a way that $|z(w_1)|<|z(w_2)|$.

\pf Let us begin with the first line of the formula \eqref{pot} for the potential $F$ represented as
\beq\label{nov1}
-\frac12 \frac1{(2\pi i)^2}\oint\oint_{|z_1| < |z_2|} \frac{w(z_1)}{z_1} \frac{w(z_2)}{z_2} \,{\rm Li}_1\left(\frac{z_1}{z_2}\right) \, dz_1\, dz_2.
\eeq
In this computation we will use the identity
\beq\label{poly1}
z\frac{d}{dz} {\rm Li}_{n+1}(z) = {\rm Li}_n(z)
\eeq
for
polylogarithms
$$
{\rm Li}_n(z) =\sum_{k\geq 1} \frac{x^k}{k^n}.
$$
Note that, in particular
\beq\label{poly2}
{\rm Li}_1(z) =-\log(1-z).
\eeq
Replacing the integration variable $z\mapsto w$ and integrating twice by parts we get, with the help of \eqref{poly1}
\eqa
&&
-\frac12 \frac1{(2\pi i)^2}\oint\oint_{|z_1| < |z_2|} \frac{w(z_1)}{z_1} \frac{w(z_2)}{z_2} \,{\rm Li}_1\left(\frac{z_1}{z_2}\right) \, dz_1\, dz_2
\nn\\
&&
=-\frac12 \frac1{(2\pi i)^2}\oint\oint_{|z(w_1)| < |z(w_2)|} w_1\,  w_2 \,{\rm Li}_1\left(\frac{z(w_1)}{z(w_2)}\right)\frac{z'(w_1)}{z(w_1)} \frac{z'(w_2)}{z(w_2)}  \, dw_1\, dw_2
\nn\\
&&
=\frac12 \frac1{(2\pi i)^2}\oint\oint_{|z(w_1)| < |z(w_2)|}{\rm Li}_3\left(\frac{z(w_1)}{z(w_2)}\right)\, dw_1\, dw_2.
\nn
\eeqa
This gives the first term in \eqref{pot1}.
Next, since
$
\bar u_0  =  v,
$
using  \eqref{kvad1} we derive that $\bar u_0 +u_0=-t_{-1}$. So we will transform the integral in the second line of \eqref{pot} as follows:
\eqa
&&
\frac1{2\pi i} \oint \frac{w(z)}{z} \log \frac{w(z)}{z} \, dz -\left( u_0 + \bar u_0\right) =-\frac1{2\pi i} \oint w \log \left(\frac{z(w)}{w}\right) \, \frac{z'(w)}{z(w)} \, dw +t_{-1}
\nn\\
&&
-\frac1{2\pi i} \oint w \,  \log \left(\frac{z(w)}{w}\right) \left[ \frac{z'(w)}{z(w)} -\frac1{w}\right] \, dw = \frac12 \frac1{2\pi i} \oint \left(  \log \frac{z(w)}{w}\right)^2 dw.
\nn
\eeqa
Multiplying by
$$
\frac12 (\bar u_0 - u_0) = v+\frac12 t_{-1}
$$
we arrive at the first term in the second line of \eqref{pot1}. The last line in \eqref{pot} we represent as follows, using $w_1=\bar u_1 + 1$ and $w_{-1}=u_{-1}+\bar u_{-1}$,
\eqa
&&
\frac12 \bar u_0^2 \log \bar u_{-1} + u_{-1}+\bar u_{-1} +\bar u_1 \bar u_{-1} = \frac12 v^2 u + w_{-1}+ e^u \bar u_1 = \frac12 v^2 u + w_{-1}+ e^uw_1 -e^u
\nn\\
&&
=\frac12 v^2 u - e^u + \frac1{2\pi i}\oint \left( \frac{e^u}{z(w)} -z(w)\right)\, dw
\nn
\eeqa
since
$$
w_1 =\frac1{2\pi i} \oint w(z) \frac{dz}{z^2} =\frac1{2\pi i} \oint w\, \frac{z'(w)}{z^2(w)} \, dw = \frac1{2\pi i} \oint \frac{dw}{z(w)}
$$
and
$$
w_{-1} =\frac1{2\pi i} \oint w(z)\, dz=\frac1{2\pi i} \oint w\, z'(w) \, dw =-\frac1{2\pi i} \oint z(w) \, dw.
$$
Putting all these terms together we arrive at \eqref{pot1}. \epf

We are now ready to complete the proof of the Main Theorem \ref{t-main}.
Define on the tangent space $T_{(\lambda, \bar\lambda)}M_0$ a symmetric trilinear form
\eqa\label{corr}
&&
<\pal_1\cdot \pal_2, \pal_3>=
\nn\\
&&
\\
&&
= \frac1{4\pi i}\oint_{|z|=1} \frac{\pal_1 w\, \pal_2 w\, \pal_3 s +\pal_1 w \,\pal_2 s\, \pal_3 w+ \pal_1s\,\pal_2  w\, \pal_3 w-s'\,\pal_1 w\, \pal_2 w\, \pal_3 w}{z^2 w'}
\, dz
\nn\\
&&
\nn\\
&&
- {\res}_{z=0} \frac{\pal_1 (\bar\lambda -\ell)
\, \pal_2 \ell\, \pal_3 \ell
+\pal_1 \ell\, \pal_2 (\bar\lambda -\ell) \,\pal_3 \ell +\pal_1 \ell \,\pal_2 \ell\, \pal_3 (\bar\lambda -\ell) 
+ \pal_1 \ell\,\pal_2 \ell\, \pal_3 \ell}{z^2 \bar\lambda'}\, dz
\nn
\eeqa
(all differentiations of the functions $w=w(z)$, $s=s(z):=\bar\lambda(z)-\lambda(z)$, $\ell=\ell(z)$, $\bar\lambda=\bar\lambda(z)$ have to be done keeping $z=\mbox{const}$). 

\begin{lemma} Let  $\pal_1$, $\pal_2$, $\pal_3$ be  flat vector fields
$\pal/\pal t_i$, $\pal/\pal u$ or $\pal/\pal v$. Then the trilinear form \eqref{corr} coincides with the triple derivatives of the potential \eqref{pot}:
\beq\label{trip0}
<\pal_1\cdot \pal_2, \pal_3>=\pal_1\pal_2\pal_3 F.
\eeq
\end{lemma}

\pf Computation of triple derivatives
of the potential \eqref{pot1} by applying \eqref{wc} is straightforward:
\eqa\label{trip}
&&
\frac{\pal^3F}{\pal t_i \pal t_j\pal t_k}=\frac12 \frac1{(2\pi i)^2} \oint_\Gamma\oint_\Gamma
\frac{z(w_1)}{z(w_2)-z(w_1)}(w_1^i-w_2^i)(w_1^j-w_2^j)(w_1^k-w_2^k) \, dw_1\, dw_2
\nn\\
&&
\\
&&
-\frac1{2\pi i} \oint_\Gamma \left( z(w) +\frac{e^u}{z(w)}\right) w^{i+j+k} dw + \frac12 \left[\delta_{i,-1} \delta_{j+k,-1} + \delta_{j,-1} \delta_{k+i, -1} +\delta_{k,-1} \delta_{i+j,-1}\right]
\nn\\
&&
\frac{\pal^3F}{\pal t_i \pal t_j\pal v} =\delta_{i+j,-1}
\nn\\
&&
\frac{\pal^3 F}{\pal v^2\pal u}=1
\nn\\
&&
\frac{\pal^3F}{\pal t_i \pal t_j\pal u} =\frac{e^u}{2\pi i} \oint_\Gamma \frac{w^{i+j}}{z(w)} dw
\nn\\
&&
\frac{\pal^3F}{\pal t_i \pal u^2}=-\frac{e^u}{2\pi i} \oint_\Gamma \frac{w^{i}}{z(w)} dw
\nn\\
&&
\frac{\pal^3F}{\pal u^3} =\frac{e^u}{2\pi i} \oint_\Gamma \frac{dw}{z(w)}  -e^u= \bar u_1 \, e^u
\nn
\eeqa
all other triple derivatives vanish.

Let us start with the first integral. We open the brackets and return to the integration in $z_1=z(w_1)$, $z_2=z(w_2)$ in order to obtain the representation
\eqa
&&
\frac12 \frac1{(2\pi i)^2} \oint_\Gamma\oint_\Gamma
\frac{z(w_1)}{z(w_2)-z(w_1)}(w_1^i-w_2^i)(w_1^j-w_2^j)(w_1^k-w_2^k) \, dw_1\, dw_2 
\nn\\
&&
\nn\\
&&
=I_1(i,j,k)+I_2(i,j,k)+I_3(i,j,k)+I_4(i,j,k)
\nn
\eeqa
where
\eqa
&&\label{int1}
I_1(i,j,k)=\frac12 \frac1{(2\pi i)^2} \oint\oint_{|z_1|<|z_2|} \frac{z_1}{z_2-z_1} w_1^{i+j+k} w_1' w_2' dz_1\, dz_2
\nn\\
&&
\nn\\
&&\label{int2}
I_2(i,j,k)=-\frac12 \frac1{(2\pi i)^2} \oint\oint_{|z_1|<|z_2|} \frac{z_1}{z_2-z_1} 
\left( w_1^{i+j} w_2^k + w_1^{j+k} w_2^i +w_1^{i+k} w_2^j\right) \, w_1' w_2' dz_1\, dz_2
\nn\\
&&
\nn\\
&&\label{int3}
I_3(i,j,k)=\frac12 \frac1{(2\pi i)^2} \oint\oint_{|z_1|<|z_2|} \frac{z_1}{z_2-z_1} 
\left( w_1^i w_2^{j+k} + w_1^j w_2^{k+i} + w_1^k w_2^{i+j}\right) \, w_1' w_2' dz_1\, dz_2
\nn\\
&&
\nn\\
&&\label{int4}
I_4(i,j,k) = -\frac12 \frac1{(2\pi i)^2} \oint\oint_{|z_1|<|z_2|} \frac{z_1}{z_2-z_1} 
w_2^{i+j+k} w_1' w_2' dz_1\, dz_2.
\nn
\eeqa
Here we denote
$$
w_1=w(z_1), \quad w_2 = w(z_2).
$$
Integrating in $z_2$ we represent the first integral in the form
$$
I_1(i,j,k)=\frac1{4\pi i} \oint_{|z|=1}  z\, w^{i+j+k} w' (w')_{\geq 0} dz.
$$
Similarly,
$$
I_2(i,j,k) =-\frac1{4\pi i} \oint_{|z|=1}  z\,w'  \left[ w^{i+j} (w^k w')_{\geq 0} + w^{j+k} (w^i w')_{\geq 0} + w^{k+i} (w^j w')_{\geq 0} \right]\, dz
$$
etc. Using the identity \eqref{ide} we rewrite
\eqa
&&
I_1(i,j,k)+I_4(i,j,k)=\frac1{4\pi i} \oint z\, w'  w^{i+j+k} (w')_{\geq 0} \,dz-\frac1{4\pi i} \oint z\, w'w^{i+j+k}\,(w')_{\leq -2}\,  dz
\nn\\
&&
\nn\\
&&
=\frac1{4\pi i} \oint z\, w'  w^{i+j+k}\,\Pi w'\, dz
\nn
\eeqa
where the operator $\Pi$ is the difference of two projectors:
\beq\label{pi}
\Pi\left( f(z)\right) =(f)_{\geq 0} -(f)_{\leq -1}.
\eeq
In a similar way we find that
\eqa
&&
I_2(i,j,k)+I_3(i,j,k)=
\nn\\
&& -\frac1{4\pi i} \oint z\,w'\left[ w^{i+j} \Pi(w^k w')
+w^{j+k} \Pi(w^i w') + w^{k+i} \Pi (w^j w')\right]\, dz
\nn\\
&&
\nn\\
&&
-\frac12 \left[ \delta_{i+j, -1}\delta_{k,-1} +\delta_{j+k,-1} \delta_{i,-1} +\delta_{k+i,-1} \delta_{j,-1}\right].
\nn
\eeqa
Thus
\eqa
&&
\frac{\pal^3F}{\pal t_i \pal t_j\pal t_k}=
\nn\\
&&
=-\frac1{4\pi i} \oint z\,w'\left[ w^{i+j} \Pi(w^k w')
+w^{j+k} \Pi(w^i w') + w^{k+i} \Pi (w^j w')- w^{i+j+k}\,\Pi w' \right]\, dz
\nn\\
&&
\nn\\
&&
-\frac1{2\pi i} \oint \left( z+\frac{e^u}{z}\right) w^{i+j+k} w'\, dz
.
\nn
\eeqa
On the other side, evaluation of the expression \eqref{corr} with
$$
\pal_1=\frac{\pal}{\pal t_i}, \quad \pal_2=\frac{\pal}{\pal t_j}, \quad \pal_3=\frac{\pal}{\pal t_k}
$$
using
$$
\frac{\pal s(z)}{\pal t_n} =-z\, \Pi (w^n w')
$$
(see \eqref{lamt}) yields
\eqa
&&
< \pal_1\cdot \pal_2, \pal_3> = -\frac1{4\pi i} \oint z\,w'\left[ w^{i+j} \Pi(w^k w')
+w^{j+k} \Pi(w^i w') + w^{k+i} \Pi (w^j w')\right]\, dz
\nn\\
&&
+\frac1{4\pi i} \oint z\,s'(z) \, w^{i+j+k} w'\, dz.
\nn
\eeqa
Since
$$
s'(z) = \Pi w' -2\left( 1+\frac{e^u}{z^2}\right)
$$
one finally obtains
\eqa
&&
< \pal_1\cdot \pal_2, \pal_3> =\frac{\pal^3 F}{\pal t_i \pal t_j \pal t_k}.
\nn
\eeqa

Next, taking
$$
\pal_1=\frac{\pal}{\pal t_i}, \quad \pal_2=\frac{\pal}{\pal t_j}, \quad \pal_3 =\frac{\pal}{\pal v}
$$
gives
$$
<\pal_1\cdot \pal_2, \pal_3> =\frac1{4\pi i}\oint \frac{\pal_1 w\, \pal_2 w\, \pal_3 s}{z^2 w'}\, dz=\frac1{2\pi i}\oint \frac{\pal_1 w\, \pal_2 w}{z^2 w'}\, dz =<\pal_1, \pal_2>=\frac{\pal^3 F}{\pal t_i \pal t_j \pal v}
$$
(we use that $\pal_v s(z) =2$). A similar computation works for $\pal_1=\pal_2= \pal/\pal u$, $\pal_3=\pal/\pal v$. For the choice
$$
\pal_1=\frac{\pal}{\pal t_i}, \quad \pal_2=\frac{\pal}{\pal t_j}, \quad \pal_3 =\frac{\pal}{\pal u}
$$
using
$$
\frac{\pal s(z)}{\pal u} =2\, \frac{e^u}{z}
$$
we obtain
$$
<\pal_1\cdot \pal_2, \pal_3> =\frac{e^u}{2\pi i}\oint \frac{w^{i+j}w'}{z}\, dz=\frac{\pal^3 F}{\pal t_i \pal t_j \pal u}.
$$
In order to perform a similar computation for
$$
\pal_1=\pal_2 =\frac{\pal}{\pal u}, 
\quad \pal_3=\frac{\pal}{\pal t_i}
$$
one has to use the second line in the formula \eqref{corr}. In this case
\eqa
&&
<\pal_1\cdot \pal_2, \pal_3> =\res_{z=0} \frac{e^{2u} (w^i w')_{\geq 0}}{z^3 \bar\lambda'}dz=e^{2u} \res\, w^i w' \, \left( \frac1{z^3 \bar \lambda'}\right)_{\leq -1}dz
\nn\\
&&
=-e^u\,\res\, \frac{w^i w'}{z}dz=\frac{\pal^3 F}{\pal u^2 \pal t_i}.
\nn
\eeqa
In the remaining cases the computation is even simpler. \epf

The next step in the proof of the Main Theorem \ref{t-main} is in the following

\begin{lemma}
The isomorphism \eqref{eta1} identifies the rank three symmetric tensor \eqref{corr} on the tangent space $TM_0$ with the one on the cotangent space given in  \eqref{inv}. 
\end{lemma}

\pf Raising the indices $i$, $j$, $k$ in the formula
\eqa
&&
\left\langle \frac{\pal}{\pal t_i} \cdot \frac{\pal}{\pal t_j}, \frac{\pal}{\pal t_k}\right\rangle=
\nn\\
&&
=-\frac1{4\pi i} \oint z\,w'\left[ w^{i+j} \Pi(w^k w')
+w^{j+k} \Pi(w^i w') + w^{k+i} \Pi (w^j w')- w^{i+j+k}\,\Pi w' \right]\, dz
\nn\\
&&
\nn\\
&&
-\frac1{2\pi i} \oint \left( z+\frac{e^u}{z}\right) w^{i+j+k} w'\, dz
\nn
\eeqa
we obtain
\eqa
&&
< dt_i\cdot dt_j, dt_k >_* =
\nn\\
&&
=-\frac1{4\pi i} \oint z\,w'\left[ w^{-i-j-2} \Pi(w^{-k-1} w')
+w^{-j-k-2} \Pi(w^{-i-1} w') + w^{-k-i-2} \Pi (w^{-j-1} w')
\right.
\nn\\
&&\left.
- w^{-i-j-k-3}\,\Pi w' \right]\, dz
-\frac1{2\pi i} \oint \left( z+\frac{e^u}{z}\right) w^{-i-j-k-3} w'\, dz.
\nn
\eeqa
We will now derive the same formula by using the multiplication \eqref{prod} and the bilinear form \eqref{eta2} on the cotangent bundle.

We will need the formula for the Jacobi matrix of the coordinate transformation $({\bf w}, u, v) \mapsto ({\bf t}, u, v)$. It is not difficult to show that
\beq\label{jac}
\frac{\pal t_n}{\pal w_m} =-\frac1{2\pi i}\oint w^{-n-1} z^{m-1} dz, \quad m, n\in \mathbb Z.
\eeq
From this we derive the following representation for the pair of functions $\hat{dt_n}\in \dot {\mathcal H}(D_0)\oplus {\mathcal H} (D_\infty) =T^*M$
representing the 1-form $dt_n =\sum_m \frac{\pal t_n}{\pal w_m}\, dw_m$:
\beq\label{jac1}
\hat{dt_n} = -\frac1{z} \left( w^{-n-1}(z)_{\geq 0}, w^{-n-1}(z)_{\leq 1}\right).
\eeq
Substitution into the formula \eqref{prod} produces the following vector in $T^*M$:
\eqa
&&
\hat{dt_i}\cdot \hat {dt_j} =
\nn\\
&&
=\left({w^{-i-1}}_{\geq 0} \left[ \lambda' {w^{-j-1}}_{\geq 0} +\bar\lambda' {w^{-j-1}}_{\leq 1}\right]_{\geq 0}
+ {w^{-j-1}}_{\geq 0} \left[ \lambda' {w^{-i-1}}_{\geq 0} +\bar\lambda' {w^{-i-1}}_{\leq 1}\right]_{\geq 0}
\right.
\nn\\
&&
-\left[ \lambda' {w^{-i-1}}_{\geq 0}\,{w^{-j-1}}_{\geq 0} +\bar\lambda'\,\left( {w^{-i-1}}_{\geq 0}\, {w^{-j-1}}_{\leq 1} +{w^{-j-1}}_{\geq 0}\, {w^{-i-1}}_{\leq 1}\right)\right]_{\geq -1},
\nn\\
&&
\nn\\
&&
-{w^{-i-1}}_{\leq 1} \left[ \lambda' {w^{-j-1}}_{\geq 0} +\bar\lambda' {w^{-j-1}}_{\leq 1}\right]_{\leq -1}
- {w^{-j-1}}_{\leq 1} \left[ \lambda' {w^{-i-1}}_{\geq 0} +\bar\lambda' {w^{-i-1}}_{\leq 1}\right]_{\leq -1}
\nn\\
&&\left.
+\left[ \bar\lambda' {w^{-i-1}}_{\leq 1}\,{w^{-j-1}}_{\leq 1} +\lambda'\,\left( {w^{-i-1}}_{\geq 0}\, {w^{-j-1}}_{\leq 1} +{w^{-j-1}}_{\geq 0}\, {w^{-i-1}}_{\leq 1}\right)\right]_{\leq 0}
\right)
\nn
\eeqa
Similarly, after substitution of \eqref{jac1} into \eqref{eta1} we obtain
\eqa
&&
\eta\left( \hat{dt_k}\right)=
-z\left( \left[\lambda' {w^{-k-1}}_{\geq 0} +\bar \lambda' {w^{-k-1}}_{\leq 1} \right]_{\leq -1} +\lambda' \,  {w^{-k-1}}_{\leq -1},
\right.
\nn\\
&&
\left.
\qquad\qquad\qquad\quad\left[ \lambda' {w^{-k-1}}_{\geq 0} +\bar \lambda' {w^{-k-1}}_{\leq 1} \right]_{\geq 0} +\bar\lambda'\,  {w^{-k-1}}_{\geq 2} 
\right).
\nn
\eeqa
Substituting
\eqa
&&
\lambda'= (w')_{\leq -2} +1+\frac{e^u}{z^2}
\nn\\
&&
\bar\lambda'= (w')_{\geq ~0} -1-\frac{e^u}{z^2}
\nn
\eeqa
after a somewhat lengthy computation we obtain that
$$
\langle \hat{dt_i}\cdot \hat{dt_j}, \eta(\hat{dt_k})\rangle= < dt_i\cdot dt_j, dt_k >_*.
$$
In a similar way we check that
\eqa
&&
\langle \hat{dt_i}\cdot \hat{dt_j}, \eta(\hat{du})\rangle= < dt_i\cdot dt_j, du >_*
\nn\\
&&
\langle \hat{du}\cdot \hat{du}, \eta(\hat{dt_k})\rangle= < du\cdot du, dt_k >_*
\nn\\
&&
\langle \hat{dt_i}\cdot \hat{dt_j}, \eta(\hat{dv})\rangle= < dt_i\cdot dt_j, dv >_*
\eeqa
etc. where
$$
\hat{du}= \left(0, e^{-u}\right), \quad \hat{dv}=\left( 0, \frac1{z}\right).
$$
The last simple step of the proof is in verifying the quasihomogeneity identity
\eqa\label{qua}
&&
E\, F=2 F + \frac12 (\bar u_0 - u_0)\frac1{2\pi i} \oint w(z) \, \frac{dz}{z} +\bar u_{0}^2 
\nn\\
&&
\qquad = 2 F-\frac12 ( v+t_{-1})t_{-1} + v^2 + \mbox{linear terms}.
\eeqa
This completes the proof of the Lemma. \epf

Let us now explain in what sense the above algebra on the cotangent space is semisimple on the open subset $M_{s\, s}\subset M_0$ 
defined by imposing an additional condition
\beq\label{ss0}
\lambda'(p)\bar\lambda''(p)-\bar\lambda'(p)\lambda''(p)\neq 0\quad \mbox{for any}\quad p\in S^1.
\eeq
For every $p\in S^1$ define a linear functional $d\mu(p)$ by the formula
\beq\label{ss1}
\langle d\mu(p), \hat\alpha\rangle= \frac{\alpha(p)}{\lambda'(p)}-\frac{\bar\alpha(p)}{\bar\lambda'(p)}.
\eeq
The functionals $d\mu(p)$ span the cotangent space to $M_{s\, s}$ (actually, at the points where $\lambda'(p)\neq 0$, $\bar\lambda'(p)\neq 0$ for $p\in S^1$; this restriction will be eliminated in a second). Indeed, the vector $\hat\alpha=(\alpha(z), \bar\alpha(z))$ can be reconstructed from knowing all the values
$$
a(p):= \langle d\mu(p), \hat\alpha\rangle \quad \mbox{for all} \quad p\in S^1
$$
by the following procedure:
\eqa\label{ss2}
&&
\alpha(z) =\quad  \lambda'(z) [a(z)]_{\leq 0}
\nn\\
&&
\\
&&
\bar\alpha(z) =- \bar\lambda'(z) [a(z)]_{\geq 1}.
\nn
\eeqa
It is convenient to change normalization of these 1-forms introducing
\beq\label{ss3}
du(p) = -\frac{\lambda'(p) \bar\lambda'(p)}{\lambda'(p)+\bar\lambda'(p)} d\mu(p) = \frac{\lambda'(p) \, d\bar\lambda(p) -\bar \lambda'(p)\, d\lambda(p)}{\lambda'(p)+\bar\lambda'(p)}.
\eeq

\begin{lemma} \label{prop-ss}The inner products and multiplication of the 1-forms $du(p)$ is given by the following expressions
\eqa\label{ss4}
&&
<du(p), du(q)>_* =f(p)\delta(p-q)
\\
&&
du(p)\cdot du(q) =f(p)\delta(p-q) \, du(p)
\label{ss5}
\\
&&
f(p)=-p^2 \frac{\lambda'(p) \bar\lambda'(p)}{\lambda'(p)+\bar\lambda'(p)} 
\eeqa
where the delta-function on the circle is defined by
$$
\delta(p-q) =\sum_{k\in\mathbb Z} \frac{p^k}{q^{k+1}}, \quad \frac1{2\pi i} \oint_{|q|=1} f(q) \delta(p-q)\, dq=f(p).
$$
\end{lemma}

\pf is given by a simple computation using \eqref{eta3}, \eqref{dlam3}. \epf

Thus, the 1-forms $du(p)$ are idempotents of the Frobenius algebra on $T^*M_{s\, s}$. The Theorem is proved. \epf

The reader familiar with the theory of finite-dimensional semisimple Frobenius manifolds certainly remembers that the basic idempotents in the {\it tangent bundle} can be represented \cite{D92} as partial derivatives along the canonical coordinates $u_1$, \dots, $u_n$.  These partial derivatives span the tangent space; they satisfy
\eqa
&&
\frac{\pal}{\pal u_i}\cdot \frac{\pal}{\pal u_j} =\delta_{ij} \frac{\pal}{\pal u_i}
\nn\\
&&
\left\langle \frac{\pal}{\pal u_i}, \frac{\pal}{\pal u_j}\right\rangle =\eta_{ii}(u)\, \delta_{ij}.
\nn
\eeqa
The canonical coordinates can be chosen in such a way that
$$
\langle du_i, E\rangle =u_i, \quad i=1, \dots, n.
$$
The multiplication and inner products of the differentials of the canonical coordinates satisfy
\eqa
&&
< du_i, du_j>_* =\eta_{ii}^{-1}(u) \delta_{ij}
\nn\\
&&
du_i \cdot du_j = \eta_{ii}^{-1}(u) \delta_{ij} du_i.
\nn
\eeqa
Using these finite-dimensional hints one arrives at the following construction of the canonical coordinates. Consider the curve $\Sigma$ defined in \eqref{sig1}.
The analytic curve $\Sigma$ is smooth, i.e. $\sigma'(p)\neq 0$ due to the assumptions \eqref{ss0}. Denote $M^0_{s\, s}\subset M_{s\, s}$ the subset consisting of pairs $(\lambda, \bar\lambda)$ such that the curve $\Sigma$ does not intersect itself. Consider the function \eqref{sig2} on the curve $\Sigma$:
$$
u_\sigma:= \left[\sigma\,\bar\lambda(p) +(\sigma-1)\, \lambda(p)\right]_{p=p(\sigma)}, \quad \sigma\in\Sigma
$$
where $p=p(\sigma)$ is the inverse map. The identity \eqref{sig3} holds true for any $\sigma\in\Sigma$. Varying the curve $\Sigma$ we obtain the variation of the point of the Frobenius manifold defined by the equation \eqref{sig3}.
 
We have to establish that the functionals \eqref{sig2} are the canonical coordinates on $M^0_{s\, s}\subset M_0$.
Indeed, taking the differential of \eqref{sig2} one obtains, due to the equation \eqref{sig3}
the 1-form \eqref{ss3}
$$
du_\sigma = du(p)_{p=p(\sigma)}.
$$ 
This proves Proposition \ref{prop4}. \epf

\begin{exam} Let us compute the Frobenius algebra structure on the two-dimensional locus $M^2_0\subset M_0$ defined by
\beq\label{locus1}
\lambda=z-v-\frac{e^u}{z}, \quad\bar\lambda=\frac{e^u}{z}+v.
\eeq
The curve $\Gamma=w(S^1)$ in this case is the unit circle with the standard parametrization. So the tangent space to $M_0$ at the points of $M^2_0$ coincides with the Cartesian product of the space of vector fields on the circle spanned by
\beq\label{locus2}
X_n = z^{n+1} \frac{\pal}{\pal z}=-\frac{\pal}{\pal t_n}, \quad n\in\mathbb Z
\eeq
and the two-dimensional space with the basis $e=\pal/\pal v$ and $\pal/\pal u$. The multiplication table of these vector fields reads
\eqa\label{locus3}
&&
X_i \cdot X_j = \frac12\left[ \theta(i)+\theta(j)+\theta(-i-j-2)+1\right]  X_{i+j+1} +\delta_{i+j,-1} \frac{\pal}{\pal u}
\nn\\
&&
\qquad\qquad +e^u \left[ X_{i+j-1} +\delta_{i+j,0} \frac{\pal}{\pal v}\right]
\\
&&
\frac{\pal}{\pal u}\cdot X_i = e^u \left[ X_{i-1} +\delta_{i,0}  \frac{\pal}{\pal v}\right]
\nn\\
&&
\frac{\pal}{\pal u}\cdot \frac{\pal}{\pal u}=e^u\, X_{-1}
\nn
\eeqa
where $\theta$ is the step function,
$$
\theta(n)=\left\{ \begin{array}{rc} 1, & n\geq 0\\
-1, & n<0\end{array}\right.
$$
\end{exam}

\begin{remark} A combination of the limit
$$
{\rm Re} \, u\to -\infty
$$
and the projector
$$
{\rm pr}: TM_0\to TM_0, \quad {\rm pr}\left( \frac{\pal}{\pal u}\right) = {\rm pr}\left( \frac{\pal}{\pal v}\right)=0, \quad {\rm pr}\left( \frac{\pal}{\pal t_i}\right)=\frac{\pal}{\pal t_i}
$$
provides the tangent planes to the space $M_{\rm red}$ of parametrized analytic curves $\{ z\mapsto w(z), ~ |z|=1\}\in M_{\rm red}$ with a structure of Frobenius algebra. The nondegenerate invariant inner product of tangent vectors is given by the first term in the formula \eqref{eta4}, the trilinear symmetric form $<\pal_1\cdot \pal_2, \pal_3>_{M_{\rm red}}$ is given by the triple derivatives of the reduced potential
\beq\label{potred}
F_{\rm red}({\bf t}) =\frac12 \frac1{(2\pi i)^2}\oint_\Gamma\oint_\Gamma {\rm Li}_3 \left( \frac{z(w_1)}{z(w_2)}\right)\, dw_1\, dw_2 -\frac1{2\pi i} \oint_\Gamma  z(w)\, dw
\eeq
For example, specializing the Frobenius algebra at the point $w(z)\equiv z$ one obtains the following graded Frobenius algebra with no unit
\eqa\label{red1}
&&
X_i \cdot X_j = \frac12\left[ \theta(i)+\theta(j)+\theta(-i-j-2)+1\right]  X_{i+j+1}
\\
&&
< X_i, X_j> =\delta_{i+j, -1}
\nn
\eeqa
(cf. \eqref{locus3} above), $\deg X_i =i+1$.
\end{remark}

\begin{exam}\label{ex215} Let us consider a two-dimensional locus
$M^2\subset M_0$ defined by the equation
$$
\lambda(z) =\bar\lambda(z).
$$
From the explicit formulae \eqref{trip} one conclude that $M^2$ is a Frobenius submanifold isomorphic to the quantum cohomology of ${\bf P}^1$
$$
\frac{\pal}{\pal u}\cdot \frac{\pal}{\pal u}= e^u \frac{\pal}{\pal v}, \quad \frac{\pal}{\pal u}\cdot \frac{\pal}{\pal v}=\frac{\pal}{\pal u}, \quad \frac{\pal}{\pal v}\cdot \frac{\pal}{\pal v}=\frac{\pal}{\pal v}, \quad \left\langle \frac{\pal}{\pal u}, \frac{\pal}{\pal v}\right\rangle=1
$$
(also describing the dispersionless limit of the standard 1+1 Toda lattice).
\end{exam}

\section{Relation with the 2D Toda hierarchy}\label{sec3}\par

We begin with the study of the intersection form of the Frobenius manifold $M_0$. 

\begin{lemma} The intersection form of the Frobenius manifold $M_0$ reads
\begin{equation} \label{g-intform1}
(\hat\omega_1 ,\hat\omega_2 )_* = \langle\hat\omega_1 , \gamma( \hat\omega_2) \rangle
\end{equation}
where the linear map $\gamma: T_{(\lm,\bar\lm)}^*M_0 \to T_{(\lm,\bar\lm)} M_0$ is defined by
\eqa \label{g-intform2}
\gamma(\hat\omega) &= z^2 \big( \lm' (\varepsilon \omega + \bar\varepsilon \bar\omega )_{\leq-2} -  \varepsilon  (\lm' \omega + \bar\lm' \bar\omega )_{\leq-2}, \\
&-\bar\lm' ( \varepsilon \omega + \bar\varepsilon \bar\omega)_{\geq-1} + \bar\varepsilon  (\lm' \omega + \bar\lm' \bar\omega )_{\geq-1} \big)
\nn
\eeqa
where $\varepsilon =\varepsilon(z)= \lm (z)  - z \lm'(z)$ and $\bar\varepsilon=\bar\varepsilon(z) = \bar\lm(z) - z \bar\lm'(z)$ are the components of the Euler vector field $E=(\varepsilon,\bar\varepsilon)$.
\end{lemma}

\begin{lemma} The linear operator $\gamma$ is invertible on the open subset of $M_0$ defined by the conditions \eqref{inter4}. The inverse operator $\hat\omega=(\omega,\bar\omega) = \gamma^{-1} (\hat\alpha), \quad \hat\alpha=(\alpha,\bar\alpha)$ reads
\beq\label{inter51}
\omega=\frac1{z^2}\left( \frac1{\lambda'}\left[ \frac{\bar\lambda'\alpha -\lambda'\bar\alpha}{\lambda\bar\lambda' -\bar\lambda\lambda'}\right]_{\geq 1}\right)_{\geq 1}, \quad \bar\omega=-\frac1{z^2}\left( \frac1{\bar\lambda'}\left[ \frac{\bar\lambda'\alpha -\lambda'\bar\alpha}{\lambda\bar\lambda' -\bar\lambda\lambda'}\right]_{\leq 0}\right)_{\leq 2}.
\eeq
\end{lemma}

\pf is similar to that of Lemma \ref{lm27}. \epf

Computing the pairing 
$$
\langle \gamma^{-1}(\hat\alpha), \hat\beta\rangle=: (\hat\alpha, \hat\beta)
$$
we obtain the expression \eqref{inter5} for the intersection form on the tangent bundle.
This completes the proof of Proposition \ref{prop2}.

\medskip

Let $\cL M$ be the loop space of maps from $S^1$ to the manifold $M$. A point in $\cL M$ is given by a pair of maps $( \lm(z,x), \bar\lm (z,x) )$. A tangent vector at a point $(\lm,\bar\lm) \in \cL M$ is naturally identified with a map from $S^1$ to $TM=\mathcal{H} (D_{\infty}) \oplus \dot{\mathcal{H}} (D_0)$ and a $1$-form with a map from $S^1$ to $T^*M = \dot{\mathcal{H}} (D_{0}) \oplus \mathcal{H} (D_{\infty})$. The pairing between a vector $\hat\alpha = ( \alpha, \bar\alpha )$ and a $1$-form $\hat\omega = ( \omega , \bar\omega )$ is the natural extension of the pairing \eqref{pair}, i.e.
\beq \label{g-looppairing}
\langle\hat\omega, \hat\alpha \rangle = \frac1{2 \pi i} \oint_{S^1} \oint_{|z|=1}  [ \alpha(z,x) \omega(z,x) + \bar\alpha (z,x) \bar\omega(z,x) ] \ dz \ dx .
\eeq

The dispersionless two-dimensional Toda hierarchy is composed of two sequences of commuting vector fields on $\cL M$, denoted by times $s_n$ and $\bar{s}_n$ for $n>0$. They are defined by the Lax equations
\eqa \label{g-todalax}
&&
\frac{\pal \lm}{\pal s_n} = \{ (\lm^n)_+ , \lm \} \qquad
\frac{\pal \bar\lm}{\pal s_n} = \{ (\lm^n)_+ , \bar\lm \}  
\nn\\
&&
\\
&&
\frac{\pal \lm}{\pal \bar{s}_n} = \{ (\bar\lm^n)_- , \lm \}  \qquad
\frac{\pal \bar\lm}{\pal \bar{s}_n} = \{ (\bar\lm^n)_- , \bar\lm \}  .
\nn 
\eeqa
The bracket of two functions of $z$ and $x$ is given by
\beq
\{ f, g \} = z \frac{\pal f}{\pal z} \frac{\pal g}{\pal x} - z \frac{\pal g}{\pal z} \frac{\pal f}{\pal x} .
\eeq
It follows from standard arguments \cite{tt} that the formulas \eqref{g-todalax} provide well-defined vector fields on $\cL M$. 

A Poisson structure on $\cL M$ is defined by a map $P_i$ from the cotangent to the tangent space of  $\cL M$ at each point of the loop space, such that the associated Poisson bracket between local functionals on $\cL M$
\beq \label{g-poissondef}
\{ F , G \}_i = \langle dF , P_i(dG) \rangle
\eeq
is skew-symmetric and satisfies the Jacobi identity. A bi-Hamiltonian formulation of the  2D Toda hierarchy has been obtained in \cite{ca1}. By taking the dispersionless limit we can easily  obtain the following result.

\begin{prop} The following formulas define pair of compatible Poisson structures, $P_1$ and $P_2$, on $\cL M$ 
\eqa \label{g-poissons}
P_1(\hat\omega) = 
&\big( -\{ \lm, (z \omega - z \bar\omega)_- \} 
+( \{ \lm , z \omega \} + \{ \bar\lm , z \bar\omega \} )_{\leq0} , 
\nn\\
& \{ \bar\lm , ( z \omega - z \bar\omega )_+ \} + ( \{ \lm , z \omega \} + \{ \bar\lm , z \bar\omega \} )_{>0} 
\big) ,
\eeqa
\eqa
P_2(\hat\omega) = 
& \big( \{ \lm , ( z \lm  \omega + z \bar\lm \bar\omega )_- \} 
- \lm ( \{ \lm , z \omega \} + \{ \bar\lm , z \bar\omega \} )_{\leq0}
+ z \lm' \varphi_x , \nn\\
&-  \{ \bar\lm , (z \bar\lm \bar\omega + z \lm \omega )_+ \}
+ \bar\lm ( \{ \lm , z \omega \} + \{ \bar\lm , z \bar\omega \} )_{>0}
+ z \bar\lm' \varphi_x \big).
\label{g-poissons1}
\eeqa
Here $\lambda'=d\lambda/dz$, $\bar\lambda'=d\bar\lambda/dz$, the function $\varphi$ is given by
\beq
\varphi = \frac1{2 \pi i} \oint_{|z|=1} \big( z\lm' \omega + z \bar\lm' \bar\omega \big) dz.
\eeq

The flows \eqref{g-todalax} are Hamiltonian with respect to both Poisson structures \eqref{g-poissons} 
\eqa \label{recurt} 
&&
\frac{\pal}{\pal s_n} \cdot = \{ \cdot , H_n \}_1 = -\{ \cdot , H_{n-1} \}_2 
\nn\\
&&
\\
&&
\frac{\pal}{\pal \bar{s}_n} \cdot = \{ \cdot , \bar{H}_n \}_1 =  ~~\{ \cdot , \bar{H}_{n-1} \}_2
\nn
\eeqa
with Hamiltonians
\beq\label{hamit}
H_n =  -\frac1{2 \pi i}\oint_{S^1} \oint_{|z|=1} \frac{\lm^{n+1}}{n+1} \frac{dz}z \ dx,\qquad 
\bar{H}_n =  -\frac1{2 \pi i}\oint_{S^1} \oint_{|z|=1} \frac{\bar\lm^{n+1}}{n+1} \frac{dz}z \ dx. 
\eeq
\end{prop}

\begin{remark} Actually we use here a slightly different bi-Hamiltonian formulation of the 2D Toda hierarchy. It differs from the one obtained in \cite{ca1} only by a few, although non-trivial, signs as one can check from the definitions above.
\end{remark}

It is possible to obtain explicit expressions for the Poisson brackets on $\cL M$. 

Let us introduce $1$-forms $d\lm(p,y)$, $d\bar\lm(p,y)$ at a point of  $\cL M$ such that 
\beq
\langle d\lm(p,y), \hat\alpha \rangle = \alpha(p,y), \qquad \langle d \bar\lm (p,y) , \hat\alpha \rangle = \hat\alpha(p,y)
\eeq
for any element $\hat\alpha=( \alpha,\bar\alpha)$ of the tangent at the same point. 
Clearly they are the differentials of the functionals on $\cL M$ that evaluate $\lm$ (  $\bar\lm$ respectively ) at a point $(p,y)$ with $y \in S^1$ and $p \in D_{\infty}$ ($D_0$ respectively). As before they can be realized as
\beq \label{g-formrepr}
d\lm(p,y) = \left( \frac{p}{z} \frac1{p-z} \de(x-y),0 \right) \qquad d \bar\lm(p,y) = \left(0, \frac{z}{p} \frac1{z-p} \de(x-y) \right) .
\eeq 
The following expressions are obtained by substitution of \eqref{g-formrepr} in \eqref{g-poissons} and \eqref{g-poissons1}. The explicit form of the first Poisson bracket is
\eqa\label{first}
&&
\{ \alpha(p,x) , \beta(q,y) \}_1 = \frac{pq}{p-q}(\epsilon(\alpha)\beta_q(q,x)-\epsilon(\beta)\alpha_p(p,x)  ) \de'(x-y) 
\nn\\
&&
\\
&&
\quad\quad \quad\quad \quad\quad \quad\quad \quad   +pq \frac{\pal}{\pal q}\left( \frac{\epsilon(\alpha)\beta_x (q,x)-\epsilon(\beta)\alpha_x(p,x) }{p-q}  \right) \de(x-y)  
\nn
\eeqa
and that of the second Poisson bracket is
\eqa
&&
\{ \alpha(p,x) , \beta(q,y) \}_2 =pq \left[ \frac{ \alpha_p \beta - \beta_q \alpha  }{p-q}+  \alpha_p \beta_q \right] \de'(x-y) 
\nn\\
&&
\\
&&
\quad\quad \quad\quad \quad\quad \quad\quad \quad +pq \left[  \frac{\pal}{\pal q} \left( \frac{\alpha_x \beta - \beta_x \alpha}{p-q} \right) + \frac{\alpha_p \beta_x - \beta_q \alpha_x}{p-q} + \alpha_p \beta_{qx}  
\right] \de(x-y) .
\nn
\eeqa
As before $\alpha$, $\beta$ can take the values $\lm$, $\bar\lm$ and by definition $\epsilon(\lm)=1$ and $\epsilon(\bar\lm)=-1$. In the right-hand side of the last formula we have assumed $\alpha=\alpha(p,x)$ and $\beta=\beta(q,x)$. The subscripts stand for the partial derivatives:
$\alpha_p=\pal\alpha(p,x)/\pal p$, $\beta_x=\pal\beta(q,x)/\pal q$ etc.
	
These are Poisson brackets of hydrodynamic type. The metrics can be read easily as coefficients of $\de'(x-y)$. In this way we arrive at

\begin{prop}
The metrics associated to the Poisson pencil \eqref{g-poissons} and \eqref{g-poissons1} of the 2D Toda hierarchy coincide with the contravariant metric \eqref{eta3} and the intersection form \eqref{inter1} of the Frobenius manifold $M_0$.
\end{prop}


\begin{remark} The map
$$
M_0 \to M_{\rm red}, \quad (\lambda(z), \bar \lambda(z))\mapsto w(z)=\lambda(z)+\bar\lambda(z)
$$
induces a map of the loop spaces
\beq\label{loo1}
\cL M_0 \to \cL M_{\rm red}.
\eeq
Let us equip the second loop space $\cL M_{\rm red}$ with the Poisson structure of the two-dimensional incompressible fluid on the two-dimensional torus $\mathbb T = \{ (x_1, x_2)\sim (x_1+2\pi m, x_2+2\pi n)\}$: 
\eqa\label{div1}
&&
\{ w(x), w(y)\} = \pal_{x_1}w(x) \delta(x_1-y_1)\delta'(x_2-y_2) -\pal_{x_2}w(x) \delta'(x_1-y_1) \delta(x_2-y_2)
\\
&&
x=(x_1, x_2), \quad y=(y_1,y_2)\in \mathbb{T}. 
\nn
\eeqa 
that is, with the Lie -- Poisson bracket on the dual space to the Lie algebra $\mathcal{V}$ of divergence-free vector fields (see, e.g., \cite{aa,nov}). The loop space $\cL M_0$ will be considered as a Poisson manifold with respect to the {\em first} Poisson bracket \eqref{first}. Then the map \eqref{loo1} is a morphism of Poisson manifolds.
\end{remark}

Indeed, the Poisson brackets \eqref{first} of the point-functionals $w(z, x)=\lambda(z,x) +\bar\lambda(z,x)$ after the substitution
$$
p=e^{i\, x_1},~ q=e^{i\, y_1}, \quad x=x_2, ~ y=y_2
$$
reduce to \eqref{div1}. The functionals $w(z,x)$ commute with $u(x)$, $v(x)$ while the brackets of these two are familiar from the Hamiltonian description of the dispersionless limit $u_{tt}=\left( e^u\right)_{xx}$ of the standard 1+1 Toda lattice:
$$
\{ u(x), v(y)\} =\delta'(x-y), \quad {\rm other ~ brackets ~ vanish}.
$$

\medskip

Let us recall \cite{icm} that with an arbitrary $n$-dimensional Frobenius manifold $M$ a system of functions
$$
\theta_{\alpha,p}(v), \quad \alpha=1, \dots, n, \quad p\geq 0, \quad v\in M
$$
is associated. In particular,
$$
\theta_{\alpha,0}=v_\alpha\equiv \eta_{\alpha, \beta} v^\beta, \quad \theta_{\alpha,1}=\frac{\pal F}{\pal v^\alpha},
$$
for $p>1$ these functions are determined from the recursion
\beq\label{recur}
\frac{\pal^2 \theta_{\alpha,p}(v)}{\pal v^\lambda \pal v^\mu} = \sum_{\nu=1}^n c_{\lambda\mu}^\nu(v) \frac{\pal \theta_{\alpha,p-1}}{\pal v^\nu}
\eeq
where all derivatives have to be taken in a system of flat coordinates.
The Hamiltonians
$$
H_{\alpha,p} =\int_{S^1} \theta_{\alpha,p+1}(v(x))\, dx
$$
commute pairwise with respect to both the Poisson structures associated with the flat pencil of metrics on $M$ (see details in \cite{icm}). These Hamiltonians satisfy certain recursion relations with respect to the bihamiltonian structure. The Hamiltonians $H_{\alpha,-1}$ are Casimirs of the first Poisson structure; the Hamiltonians $H_{\alpha,0}$ generate the primary flows \eqref{inter2}: 
$$
\left\{ ~ \cdot ~, H_{\alpha,0}\right\}_1 =\frac{\pal}{\pal t^{\alpha,0}}, \quad \alpha=1, \dots, n.
$$
Let us describe the primary flows for the infinite-dimensional Frobenius manifold $M_0$.

\begin{lemma}
For the Frobenius manifold $M_0$ the primary Hamiltonians
\eqa
&&
H_{\alpha,0}=\int_{S^1} \frac{\pal F}{\pal t_\alpha}\, dx, \quad \alpha\in \mathbb Z
\nn\\
&&
H_{u,0}= \int_{S^1} \frac{\pal F}{\pal u}\, dx, \quad H_{v,0}= \int_{S^1} \frac{\pal F}{\pal v}\, dx
\nn
\eeqa
generate the equations \eqref{inter3}.
\end{lemma}

\pf We prove the theorem for the flows $\frac{\pal}{\pal t^{\alpha,0}}$, $\alpha\in\mathbb Z$ leaving the remaining cases $\frac{\pal}{\pal t^{u,0}}$ and $\frac{\pal}{\pal t^{v,0}}$ as an easy exercise  for the reader. We will first compute the derivatives $\frac{\pal}{\pal t^{\alpha,0}}$ of the variables $w(z),u,v$ and then show that they coincide with the Lax flows \eqref{inter3}.

{\bf Step 1.}  We compute $\frac{\pal z(w)}{\pal t^{\alpha,0}}$. 
\beq
\frac{\pal z(w)}{\pal t^{\alpha,0}} = \sum_i z(w) w^i \frac{\pal t_{i}}{\pal t^{\alpha,0}} = z(w) w^i 
\left[ c_{-(i+1),\alpha,\beta} t^{\beta}_x + c_{-(i+1),\alpha,u} u_x + c_{-(i+1),\alpha,v} v_x \right]\label{prim1}
\eeq
Before plugging the triple derivatives $c_{\alpha,\beta,\gamma}$ of the potential \eqref{trip} in \eqref{prim1}, we observe that they can be rewritten as follows:
\eqa
c_{\alpha,\beta,\gamma}&=&\frac1{2 \pi i}\oint \frac{z(a)a^\alpha-z(b)b^\alpha}{z(a)-z(b)} a^\gamma b^\beta da\ db +\\ 
&&+\frac1{2 \pi i}\oint \frac{z(a)}{z(a)-z(b)}a^{\beta+\gamma}b^{\alpha} da\ db -\\
&&-\frac1{2 \pi i}\oint \left(z(a)+\frac{e^u}a+\oint\frac{z(a)}{z(a)-z(r)}dr \right) a^{\alpha+\beta+\gamma} da
\eeqa
Substituing this expression into \eqref{prim1} we get:
\eqa
\frac{\pal z(w)}{\pal t^{\alpha,0}}&=&z(w)w^{\alpha}\oint\frac{1}{z(w)-z(b)}\frac{z(w)}{z(b)}z_x(b)db-
z(w)\oint\frac{b^\alpha}{z(w)-z(b)}z_x(b)db+\nn\\
&&
\label{prim:z(w)}\\
&&+z(w)z_x(w)\oint\frac{b^\alpha}{z(w)-z(b)}db-\nn\\
&&
\nn\\
&&-\left( z(w)+\frac{e^u}{z}(w)+\oint\frac{z(w)}{z(w)-z(r)}dr\right)z_x(w)w^\alpha+\nn\\
&&
\nn\\
&&+e^uu_xw^\alpha+z(w)w^\alpha v_x\nn
\eeqa
Note that this computation holds for every $\alpha\in \mathbb Z$. 

Using simple identities
\eqa
\frac{\pal w(z)}{\pal t^{\alpha,0}}&=&-w'(z)\frac{\pal z(w)}{\pal t^{\alpha,0}}(w(z))\label{id:t}\\
&&
\nn\\
w_x(z)&=&-w'(z)z_x(w(z))\label{id:x} 
\eeqa
along with \eqref{prim:z(w)} we get
\eqa
\frac{\pal w(z)}{\pal t^{\alpha,0}}&=&-w'(z)w^\alpha(z)z\oint\frac{1}{z-z(b)}\frac{z}{z(b)}z_x(b)db+
z w'(z)\oint\frac{b^\alpha}{z-z(b)}z_x(b)db+\nn\\
&&
\nn\\
&&+z w_x(z)\oint\frac{b^\alpha}{z-z(b)}db-
w_x(z) w^\alpha(z) \left(z+\frac{e^u}{z}+\oint\frac{z}{z-z(r)}dr \right)-\nn\\
&&
\nn\\
&&-e^u u_x w'(z) w^\alpha(z)-z w'z w^\alpha(z) v_x
\eeqa
Perform the change of variables $a=z(b)$ in the integrals, using \eqref{id:x} for the first two terms:
\eqa
\frac{\pal w(z)}{\pal t^{\alpha,0}}&=&w'(z)w^\alpha(z)z\oint\frac{1}{z-a}\frac{z}{a}w_x(a)da-
z w'(z)\oint\frac{w^\alpha(a)}{z-a}w_x(a)db+\nn\\
&&
\nn\\
&&+z w_x(z)\oint\frac{w^\alpha(a)}{z-a}w'(a)da-w_x(z)w^\alpha(z)\oint\frac{z}{z-a}w'(a)da-\nn\\
&&
\nn\\
&&-w_x(z) w^\alpha(z) \left(z+\frac{e^u}{z}\right)-\nn\\
&&
\nn\\
&&-e^u u_x w'(z) w^\alpha(z)-z w'z w^\alpha(z) v_x
\eeqa
All this integrals are projections of the type \eqref{proj2}. Computing them explicitly we finally get:
\eqa
\frac{\pal w(z)}{\pal t^{\alpha,0}} &=& w^{\alpha}\left\{ w(z), w(z)_{\leq 0} \right\}+
w^{\alpha} \left\{ w^{\alpha+1}(z), z-v-\scriptstyle{\frac{e^u}z} \right\}+\label{prim:w}\\ 
&&
+\left(z w'(z) w^{\alpha}(z)\right)_{<0}w_x(z) - z w'(z)\left( w^{\alpha}(z)w_x(z)\right)_{<0}\nn
\eeqa
The primary time derivatives of the coordinates $u,v$ are given directly by \eqref{trip}:
\eqa
\frac{\pal v}{\pal t^{\alpha,0}}&=&e^u \pal_x \left(-\oint\frac{w^\alpha}{z(w)}dw\right)+e^uu_x \left(-\oint\frac{w^\alpha}{z(w)}dw\right)\label{prim:v}\\
&&
\nn\\
\frac{\pal u}{\pal t^{\alpha,0}}&=&\pal_x\left(t^{-(\alpha+1)}\right)\label{prim:u}
\eeqa

{\bf Step 2.} We write the Lax flows $\frac{\pal}{\pal t^{\alpha,0}}$ for $\alpha\neq -1$ in the $w(z),u,v$, coordinates and show that they coincide with \eqref{prim:w},\eqref{prim:v} and \eqref{prim:u}. Plugging formulas \eqref{dubl1} into \eqref{inter3} we get:
\eqa
\frac{\pal w(z)}{\pal t^{\alpha,0}} &=&\frac1{\alpha+1} \left\{ w^{\alpha+1}(z)_{<0}, w(z)_{\leq 0} \right\}-
\frac1{\alpha+1} \left\{ w^{\alpha+1}(z)_{\geq 0}, w_{>0} \right\}+\label{Lax:w}\\
&&+\frac1{\alpha+1} \left\{ w^{\alpha+1}(z), z-v-\frac{e^u}z \right\}\nn\\
&&
\nn\\
\frac{\pal v}{\pal t^{\alpha,0}} &=& \left\{ \left(-\frac{w^{\alpha+1}(z)}{\alpha+1}\right)_1z, \frac{e^u}z \right\}\label{Lax:v}\\
&&
\nn\\
\frac{e^u}z\frac{\pal u}{\pal t^{\alpha,0}} &=& \left\{ \left(-\frac{w^{\alpha+1}(z)}{\alpha+1}\right)_0, \frac{e^u}z \right\}\label{Lax:u}
\eeqa
Here as above $(f(z,x))_n$ is the $n$-th coefficient of the Laurent expansion of $f(z,x)$ in the $z$ varible. Adding 
$$
\frac1{\alpha+1} \left\{ w^{\alpha+1}(z)_{\geq 0}, w(z)_{\leq 0} \right\}- \frac1{\alpha+1} \left\{ w^{\alpha+1}(z)_{\geq 0}, w(z)_{\leq 0} \right\}=0
$$ 
to \eqref{Lax:w} we get:
\eqa
\frac{\pal w(z)}{\pal t^{\alpha,0}} &=&\frac1{\alpha+1} \left\{ w^{\alpha+1}(z), w(z)_{\leq 0} \right\}+
\frac1{\alpha+1} \left\{ w^{\alpha+1}(z)_{<0}, w(z) \right\}+\\
&&+\frac1{\alpha+1} \left\{ w^{\alpha+1}(z), z-v-\frac{e^u}z \right\}\nn
\eeqa
Clearly $z\pal_z\left(f(z)\right)_{<0}=\left( z\pal_z f(z) \right)_{<0}$, $\pal_x\left(f(z)\right)_{<0}=\left( \pal_x f(z) \right)_{<0}$, hence we can rewrite the flow in the final form: 
\eqa
\frac{\pal w(z)}{\pal t^{\alpha,0}} &=& w^{\alpha}\left\{ w(z), w(z)_{\leq 0} \right\}+
w^{\alpha} \left\{ w^{\alpha+1}(z), z-v-\frac{e^u}z \right\}+\label{Lax:w:gen}\\ 
&&
+\left(z w'(z) w^{\alpha}(z)\right)_{<0}w_x(z) - z w'(z)\left( w^{\alpha}(z)w_x(z)\right)_{<0}\nn
\eeqa
The formula \eqref{Lax:w:gen} can be proven in a similar manner also for the exceptional Lax flow $\frac{\pal}{\pal t^{-1,0}}$. This formula coincides with primary flow evaluation over $w(z)$ (formula \eqref{prim:w}).   

To conclude the proof we rewrite \eqref{Lax:v} and \eqref{Lax:u} in the more convenient form:
\eqa
\frac{\pal v}{\pal t^{\alpha,0}} &=& \left(-\frac{w^{\alpha+1}(z)}{\alpha+1}\right)_1 e^u u_x + e^u \pal_x \left(-\frac{w^{\alpha+1}(z)}{\alpha+1}\right)_1\label{Lax:v:gen}\\
&&
\nn\\
\frac{\pal u}{\pal t^{\alpha,0}} &=& \pal_x \left(-\frac{w^{\alpha+1}(z)}{\alpha+1}\right)_0 \label{Lax:u:gen}
\eeqa
Also this formulas have an analog in the $\alpha=-1$ case, one just has to replace the function $\left(-\frac{w^{\alpha+1}(z)}{\alpha+1}\right)$ with $\left(-\frac{\log w(z)}{z}\right)$.

One can easily see that $\left(-\oint\frac{w^\alpha}{z(w)}dw\right)=\left(-\frac{w^{\alpha+1}(z)}{\alpha+1}\right)_1$ for $\alpha\neq-1$, while $\left(-\oint\frac{w^{-1}}{z(w)}dw\right)=\left(-\frac{\log w(z)}{z}\right)_1$. Hence \eqref{Lax:v:gen} coincides with the evaluation of the primary flow over $v$ (formula \eqref{prim:v}). Proving that \eqref{Lax:u:gen} and \eqref{prim:u} coincide only uses the definition of the flat coordinates $t_i$ in terms of $w(z)$. \epf

\begin{lemma} The Hamiltonians of the Principal Hierarchy associated with the Frobenius manifold $M_0$ satisfy the following recursion
\eqa\label{recurb}
&&
\left\{ ~\cdot ~, H_{\alpha,p-1}\right\}_2 = (p+\alpha+1) \, \left\{ ~\cdot ~, H_{\alpha,p}\right\}_1
\\
&&
\left\{ ~\cdot ~, H_{u,p-1}\right\}_2=(p+1)\, \left\{ ~\cdot ~, H_{u,p}\right\}_1
\nn\\
&&
\left\{ ~\cdot ~, H_{v,p-1}\right\}_2= p\, \left\{ ~\cdot ~, H_{v,p}\right\}_1+2\, \left\{ ~\cdot ~, H_{u,p-1}\right\}_1.
\nn
\eeqa
\end{lemma}

\pf follows from the standard formalism of the theory of Frobenius manifolds \cite{normal,icm} taking into account the quasihomogeneity degrees of the flat coordinates
$$
\deg t_\alpha =-(\alpha+1), \quad \deg v=1, \quad \deg u=0, ~ \deg e^u=2.
$$\epf

\begin{corollary} The Hamiltonians \eqref{hamit} of the dispersionless 2D Toda hierarchy commute, with respect to both the Poisson brackets with all Hamiltonians of the Principal Hierarchy.
\end{corollary}

\pf Let us prove that $\left\{ H_n, H_{\alpha,p}\right\}_1=0$. Using recursions \eqref{recurt} and \eqref{recurb} we obtain
$$
\left\{ H_n, H_{\alpha,p}\right\}_1= -\left\{ H_{n-1}, H_{\alpha,p}\right\}_2
=-(p+\alpha+2)\, \left\{ H_{n-1}, H_{\alpha,p+1}\right\}_1.
$$
Iterating we arrive at the equation
$$
\left\{ H_n, H_{\alpha,p}\right\}_1=\mbox{const} \left\{ H_{-1}, H_{\alpha,q}\right\}_1
$$
for some constant coefficient and some $q>p$. The Poisson bracket in the right-hand side vanishes since
$$
H_{-1} =-\int_{S^1}( t_{-1}+v)dx
$$
(see \eqref{kvad1}) is a Casimir of the first Poisson bracket. Similarly, using
$$
\bar H_{-1} =\int_{S^1} v\, dx
$$
we prove that 
$$
\left\{\bar H_n, H_{\alpha,p}\right\}_1=0.
$$
Commutativity of the Hamiltonians $H_n$, $\bar H_n$ with other Hamiltonians of the principal hierarchy with respect to both Poisson brackets can be proved in a similar manner. This completes the proof of the Corollary and, thus, the proof of Theorem \ref{t-main2}. \epf

In the conclusion of this Section we derive the diagonal form \eqref{riem} of the primary flows using the canonical coordinates \eqref{sig2}.
By the general definition \cite{icm} the primary time derivative of an arbitrary function $f$ on $M_0$ can be written in the form \cite{icm}
$$
\frac{\pal f}{\pal t^{i,0}}  = \frac{\pal}{\pal t_i} \cdot \frac{\pal f}{\pal x}
$$
where the product of tangent vectors $\pal/\pal t_i$ and $\pal f/\pal x$ has to be computed in the right hand side. The operator of multiplication by $\pal/\pal t_i$ becomes diagonal in the canonical coordinates with the eigenvalues
$$
\langle du(p), \frac{\pal}{\pal t_i}\rangle.
$$
Computation of this pairing with the help of \eqref{lamt} gives the needed expression. In a similar way, using \eqref{lamuv} we prove the second formula in \eqref{riem}. This completes the proof of Proposition \ref{prop3}. \epf

\begin{remark} One can also easily obtain the diagonal form of the dispersionless Toda equations:
\eqa\label{r-toda}
&&
\frac{\pal u_\sigma}{\pal s_n} = C_n(\sigma) \frac{\pal u_\sigma}{\pal x}
\nn\\
&&
\frac{\pal u_\sigma}{\pal \bar s_n} = \bar C_n(\sigma) \frac{\pal u_\sigma}{\pal x}
\nn\\
&&
C_n(\sigma) =\left[ \left(p \, \lambda'(p)\right)_{\geq 0}\right]_{p=p(\sigma)}, \quad \bar C_n(\sigma) =\left[ \left(p \, \bar\lambda'(p)\right)_{< 0}\right]_{p=p(\sigma)}
\\
&&
n=1, 2, \dots, \quad \sigma\in\Sigma.
\nn
\eeqa
\end{remark}

\setcounter{equation}{0}
\setcounter{theorem}{0}
\section{Concluding remarks}\par

In the present paper we have introduced a structure of an infinite-dimensional semisimple Frobenius manifold on the space of pairs of symbols of Lax operators of the 2D Toda hierarchy provided validity of certain analyticity conditions for the symbols. We have demonstrated that the rich geometry known from the finite-dimensional theory extends to the infinite-dimensional case. The analytic conditions for the symbols were crucial in establishing the main properties of these geometrical structures.

It would be certainly of interest to study generalizations of the above constructions to a wider class of symbols with less restrictive analytic assumptions. We plan to do it in subsequent publications.

In a subsequent publication we will also study the properties  of solutions to the Principal Hierarchy and their tau-functions. In particular we will be interested in the so-called {\it topological solution} obtained by extending the potential $F$ to the descendent time variables. The tau-function of this solution is specified by the string equation  
\eqa\label{string}
&&
\frac{\pal \log\tau}{\pal t^{v,0}}=\sum_{\alpha\in\mathbb Z} \sum_{k\geq 1} t^{\alpha,k}\frac{\pal \log\tau}{\pal t^{\alpha, k-1}} + \sum_{k\geq 1} t^{u,k}\frac{\pal \log\tau}{\pal t^{u, k-1}}+ \sum_{k\geq 1} t^{v,k}\frac{\pal \log\tau}{\pal t^{v, k-1}}
\\
&& 
\qquad\qquad + \frac12 \sum_{i+j=-1} t^{i,0}t^{j,0}+t^{u,0} t^{v,0}.
\nn
\eeqa
Note that the infinite sum in the right hand side converges on the small phase space
$$
t^{i,0} = t_i, \quad t^{u,0}=u, \quad t^{v,0}=v, \quad \mbox{other times vanish}
$$
due to \eqref{kvad}.

For more general solutions we plan to study their local structure near the singular points. For solutions to 1+1 systems the singular points of a generic solution correspond to the gradient catastrophe of a single Riemann invariant (or of a pair of complex conjugate invariants for the case of nonlinear elliptic systems), see \cite{ams}. For the solutions to the Principal Hierarchy (and, in particular, for solutions to the dispersionless 2D Toda) and other systems with continuous families of Riemann invariants one might expect a somewhat more complicated behaviour associated with appearance of singularities on the curve $\Sigma$.

The constructions of this paper can be generalized to the more general Frobenius manifolds
$M_0^{m,n}$ consisted from the pairs of functions $\lambda(z)$, $\bar\lambda(z)$ with the poles at $\infty/0$ resp. of the orders $m$ and $n$. The details will be given elsewhere. The Frobenius manifold $M_0^{m,n}$ contains a finite-dimensional submanifold (cf. Example \ref{ex215} above) associated with the so-called bigraded Toda lattice \cite{ca2}. It would be also interesting to find an extension of the Frobenius manifold $M_0$ in order to include the dispersionless limit of a more complicated reduction of the 2D Toda hierarchy constructed by E.Getzler \cite{ge} in the theory of equivariant Gromov -- Witten invariants of ${\bf P}^1$ (see also \cite{fs,mt}).

The problem of classification of infinite-dimensional semisimple Frobenius manifolds looks to be attractive. Recall \cite{D92,icm} that in  the $n$-dimensional case semisimple Frobenius manifolds are parametrized by the $n\times n$ Stokes matrices of a certain linear differential operator of order $n$ with rational coefficients.

Last but not least: we are confident that the ideas and methods of the present paper can be generalized to other 2+1 integrable systems. This will be done in subsequent publications.

\begin{acknowledgements}
 This work is
partially supported by European Science Foundation Programme ``Methods of
Integrable Systems, Geometry, Applied Mathematics" (MISGAM), Marie Curie RTN ``European Network in Geometry, Mathematical Physics and Applications"  (ENIGMA), 
and by Italian Ministry of Education, Universities and Researches (MIUR) research grant PRIN 2006
``Geometric methods in the theory of nonlinear waves and their applications". The research of one of the authors (B.D.) is partially supported by the European Research Council Advanced Grant  ``Frobenius Manifolds and Hamiltonian Partial Differential Equations (FroM-PDE). The authors thank the anonymous referee for careful reading the manuscript and correcting few mistakes.
\end{acknowledgements}

\end{document}